\documentclass[11pt]{article}
\usepackage{color} 

\usepackage{amsmath}
\usepackage{amsbsy}
\usepackage{amscd}
\usepackage{amsopn}
\usepackage{amstext}
\usepackage{amsxtra}
\usepackage{graphicx}
\usepackage{epsfig}
\usepackage{authblk}
\usepackage{cite}
\usepackage{url}
\usepackage{hyperref}

\usepackage{amsmath,amstext,amsgen,amsfonts}

\def\t{\tau}

\begin{document}
\title{{ The effects of network structure, competition and memory time on social spreading phenomena}}
\author{James P. Gleeson}
\author{ Kevin P. O'Sullivan}
\affil{MACSI, Department of Mathematics and Statistics, University of Limerick, Ireland}
\author{Raquel A. Ba\~{n}os}
\affil{Instituto de Biocomputaci{\'o}n y F{\'i}sica de Sistemas Complejos (BIFI), Universidad de Zaragoza, Mariano Esquillor s/n, 50018 Zaragoza, Spain}
\author{Yamir Moreno$^{2,}$}
\affil{Department of Theoretical Physics, Faculty of Sciences, University of Zaragoza, Zaragoza 50009, Spain}
\affil{Institute for Scientific Interchange (ISI), Turin, Italy}

\date{30 Mar 2016}

\maketitle

\begin{abstract}
Online social media have greatly affected the way in which we communicate with each other. However, little is known about what are the fundamental mechanisms driving dynamical information flow in online social systems. Here, we introduce a generative model for online sharing behavior
that is analytically tractable and which can  reproduce several characteristics of empirical micro-blogging data on hashtag usage, such as (time-dependent) heavy-tailed distributions of meme popularity. The presented framework constitutes a null model for social spreading phenomena which, in contrast to purely empirical studies or simulation-based models, clearly distinguishes the roles of two distinct factors affecting meme popularity:  the memory time of users and the connectivity structure of the social network.
\end{abstract}

\section{Introduction}

Recent advances in communication technologies and the emergence of social media have made it possible to communicate rapidly on a global scale. However, since we receive pieces of information from multiple sources, this has also made the information ecosystem highly competitive: in fact, users' influence and visibility are highly heterogeneous and  topics strive for users' attention in online social systems. Although several studies have described the dynamics of information flow in popular communication media \cite{Bakshy11,Lerman12,Weng12,Cheng14,GleesonPNAS14}, the main factors determining the observed patterns have not been identified and there is no theoretical framework that addresses this challenge. Indeed, given the potential for applications---e.g., having more efficient systems to spread information for safety and preparedness in the face of threats---a better understanding of how memes (ideas, hashtags, etc.) emerge and compete in online social networks is critical.

{
Information often spreads through a social network as a cascade: a person adopts a new behavior or installs a new app, or sends a news item or rumour to their friends (e.g., by tweeting it on Twitter). The avalanche spreads if the friends decide to also adopt the new behavior, and in turn pass on the social influence effect to their own friends, who may further propagate the behavior. Following the usage in the review \cite{Castellano09}, we apply the term ``social spreading phenomena'' to describe such cascading or ``viral'' propagation \cite{Leskovec07}. The latter term is used because the description of information spreading bears some similarity to epidemics of contagious disease; the effects of network structure on disease contagion have been well-studied by physicists \cite{PastorSatorras01}, see \cite{diseasereview} for a recent review. However, unlike epidemics of a single disease strain, we focus on social spreading phenomena that occur in the presence of competition between a large number of different items of similar type. Examples of the types of items include URLs on Twitter \cite{Bakshy11,Lerman12}, apps on Facebook \cite{Onnela10,GleesonPNAS14}, or videos on YouTube \cite{Miotto}. In each of these examples, users make choices---often influenced by the choices they have seen their friends make---and the accumulation of many choices leads to a distribution of popularity of the items: some items become extremely popular, while other items remain obscure.

To enable a succinct general description, we will call such items by Dawkins' term \cite{Dawkins}``memes'' because they are all ``elements of a culture or system of behavior passed from one individual to another by imitation...''\cite{memedefinition}. Note that we do not restrict our study only to very popular memes; indeed our interest is in understanding the entire popularity distributions of memes, from the unpopular to the very popular. This definition of a meme has also been used by researchers studying cascades on Facebook \cite{Adamic14}, the spreading of news through blogs \cite{Leskovec09}, and the popularity of hashtags on Twitter \cite{Weng12,Weng13}, but it can also be applied to analyze popularity distributions of offline items (where copying promotes spreading) such as baby names \cite{Bentley04}, dog breeds \cite{Herzog04}, and even to citations (which are a type of popularity measure) of scientific papers \cite{Simkin07,Redner98}. The memes in these examples are all relatively simple units of information that are easily identified in data sets; recent work has also demonstrated that more complex memes (represented by the appearance of common phrases, such as ``quantum'' or ``graphene'', in the scientific literature) can be recognized by their inheritance patterns in the citation network \cite{Kuhn14}.

A notable characteristic of many meme popularity distributions is that they are very fat-tailed: if a power-law distribution is fitted to the data then the power-law exponent $\tau$ is typically between 1.5 and 2, which lies outside the range of exponents produced by models of cumulative-advantage \cite{Price76,Newman05,Perc14} or preferential-attachment \cite{Barabasi99} type.
 The statistical physics of avalanches has  been studied in the context of condensed-matter systems, where the flip of a single magnetic spin domain can cause its neighboring domains to also flip and so initiate a cascade \cite{Sethna01}. If the physical parameters of such a system are tuned to place it at a critical point \cite{Stanleybook} the sizes of avalanches are power-law distributed; the sandpile model of self-organized criticality (SOC) self-tunes so that the system balances at the critical point \cite{Bakbook}. However, unlike the memoryless particles or magnetic spins that constitute the microscopic entities in condensed-matter avalanches, humans absorb and transmit information on a wide variety of timescales that range from seconds to weeks \cite{Barabasi05,Malmgren08}. Models of social interaction must therefore include ``memory'' effects (non-Markovian aspects) that lead to the emergence of characteristics that are qualitatively different from those seen in condensed-matter avalanches. The non-Markovian aspects of human temporal behavior have attracted considerable recent attention (e.g. \cite{Karsai11,Delvenne15,JoPRX14,Iribarren09,Iribarren11}), but we wish to investigate the effects  of memory on popularity avalanches caused by users choosing between multiple items that they have seen in the past.

To address this problem, we develop a theoretical framework that models how users choose among multiple sources of incoming information and affect the spreading of memes on a directed social network, like Twitter \cite{Bakshy11,Lerman12,Weng12}. Our probabilistic model, in contrast to other studies \cite{Weng12,Cheng14,GleesonPNAS14,Bentley11,Simkin07} that use intensive computational simulations to fit to data, allows us to get analytical insights into the respective roles of the network degree distribution, the memory-time distribution of users, and the  competition between memes for the limited resource of user attention.
The model is a ``null model'' in the sense that it is analytically tractable, yet realistic enough to be fitted to  empirical data and to reproduce some important characteristics of the data.
We show that fitting to time-dependent data requires a non-trivial memory-time distribution, which is not possible with the toy model  of Ref.~\cite{GleesonPRL14}, where users can remember only one meme. However, the phenomenon of ``competition-induced criticality'' that was first identified in \cite{GleesonPRL14} is shown to be robust to the inclusion of memory-times, heterogeneous user activity rates and  complex network structures in the more realistic model used here. The current model requires more sophisticated mathematical analysis than that of Ref.~\cite{GleesonPRL14} to deal with the long memory of users, but it enables us to understand how  heavy-tailed distributions of meme popularity evolve over a range of timescales, as a few memes ``go viral'' but the majority become only moderately popular.

We phrase the model in terms of meme propagation on a directed social network (like Twitter) and interpret a ``meme'' to be any distinct piece of information that is easily copied and transmitted (e.g., a hashtag or URL within a tweet). However, it should be clear that the model and its results can also be extended to the other examples of viral phenomena discussed above. For the adoption of apps on Facebook, for example, the memes are the notifications sent when a user installs an app \cite{Onnela10}. If a friend is prompted by this notification to also install the app, then the meme propagates on the network and its popularity is measured by the number of installations of the app. We show that the crucial property of the model that poises the system at criticality is the competitive pressure for the limited resource of user attention, and this property is common to a broad range of social spreading phenomena that are characterized by the availability of large time-dependent data sets.
}

The remainder of the paper is structured as follows.
The model is introduced in Sec.~\ref{model}; in Secs.~\ref{derivation} and \ref{analysis} we derive and analyze a branching-process description of the model dynamics. We confirm the results of this analysis using numerical simulations in Sec.~\ref{resultssynthetic} and then use the analytical results to fit
the model to hashtag popularities extracted from micro-blogging data in Sec.~\ref{resultsdata}, and to explain novel features of the time-dependent data. { In Sec.~\ref{limitations} we discuss limitations of the model and possible extensions of it.} Note that the Secs.~\ref{derivation} and \ref{analysis} may be omitted on a first reading without affecting the understanding of the model and the main results.

\section{Model} \label{model}

In online communication platforms like Twitter, users follow (receive the broadcasts or ``tweets'' of) other users. In graph-theoretical terms, these relationships constitute directed links from the followed node (user) to the follower (Fig.~1).
The network structure is defined  by the joint probability $p_{jk}$ that a randomly-chosen node (user) has in-degree $j$ (i.e., follows $j$ other Twitter users) and out-degree $k$ (i.e., has $k$ followers), but the network is otherwise assumed to be maximally random (a configuration model directed network). The mean degree of the network is $z=\sum_{j,k} k p_{jk} = \sum_{j,k} j p_{jk}$. If we simplify the model by assuming that all users follow $z$ others---as we sometimes do to highlight the role of the out-degree distribution---then $p_{j k}$ can be replaced with $\delta_{j,z} p_k$, where $\delta_{j,z}$ is the Kronecker delta and $p_k$ is the out-degree distribution.

\begin{figure}
\centering
\epsfig{figure=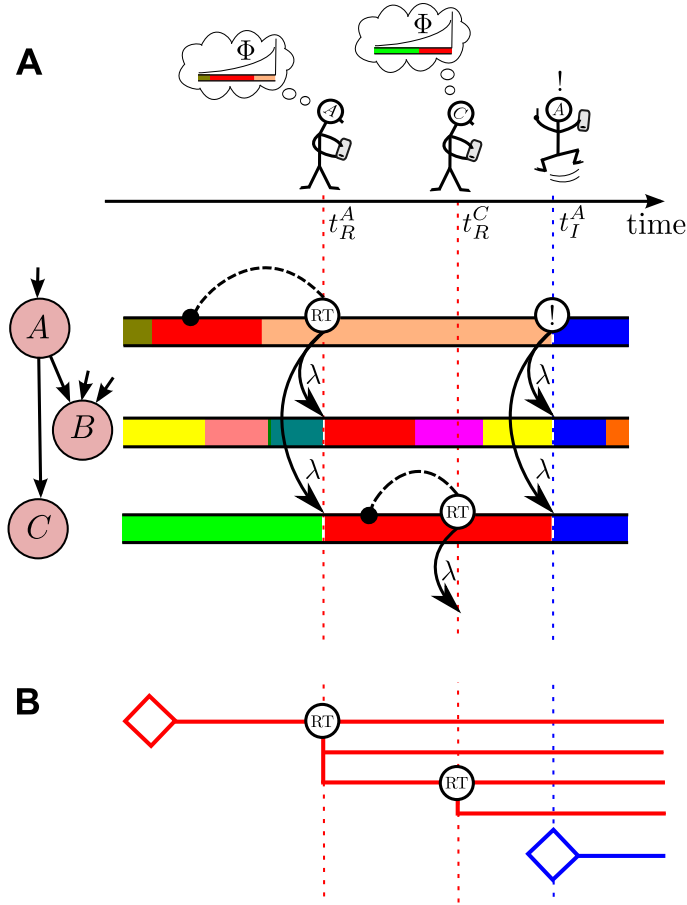,width=9.0 cm}
\caption{{Schematic of the model.} {(A)} Timeline of users' actions in a typical realization of the model. User $A$ is followed by users $B$ and $C$; arrows between nodes denote the direction of information transmission. Note that user $B$ also follows many other users, and so his stream contains more memes than the streams of $A$ or $C$.  At time $t_R^A$, user $A$ retweets a previously-seen meme (with probability $1-\mu$, given $A$ is active). She chooses the red meme to retweet, by looking backwards in her stream a distance determined by the memory-time distribution $\Phi$ (only memes that $A$ deemed ``interesting'' are shown in her stream). Her retweet of the red meme is accepted as ``interesting'' (and so inserted into their stream) by each follower of $A$ with probability $\lambda$. At time $t_R^C$ user $C$ retweets the red meme to his followers, so further increasing the popularity of the red meme. At time $t_I^A$ user $A$ innovates (a probability $\mu$ event, given $A$ is active) by inventing the new blue meme and broadcasting it to her followers. {(B)} Branching process representation (Sec.~\ref{derivation}) of the popularities of the red meme and of the blue meme. Each retweet generates new branches of the process, as the meme is inserted into the streams of followers of the tweeting user.}
\end{figure}

Each user has a ``stream'' that records all tweets received by the user, time-stamped by their arrival time. We assume that only a fraction $\lambda$ of the tweets received are deemed ``interesting'' by the user, and only the interesting tweets are considered for possible retweeting by that user.
(Here we use the term ``retweeting'' in a general sense, to include any reuse of a previously-received meme such as a hashtag: note that a meme may be retweeted more than once by a user, unlike the model of Ref.~\cite{Iribarren11}).
The activity rate of a user---the average number of tweets that she sends per unit time, i.e., the rate of the Poisson process that describes her tweeting activity---can depend on how well-connected the user is within the social network \cite{Weng12}, and we assume it depends on her in-degree $j$ and out-degree $k$ (her ``$(j,k)$-class'' for short); this assumption is supported by empirical evidence from Twitter, see Fig.~6 of Ref.~\cite{Hodas13}. The user activity rates $\beta_{jk}$ give the relative activity levels of users in the $(j,k)$ class; the rates are normalized by choosing time units so that $\sum_{j k} \beta_{j k} p_{j k}=1$. If there are $N$ users in the network, this rate implies that an average of $N$ tweets are sent in each model time unit. To simplify the analysis, we will sometimes specialize to the case where all user activity rates are equal: $\beta_{j k}=1$.

\begin{figure}
\centering
\epsfig{figure=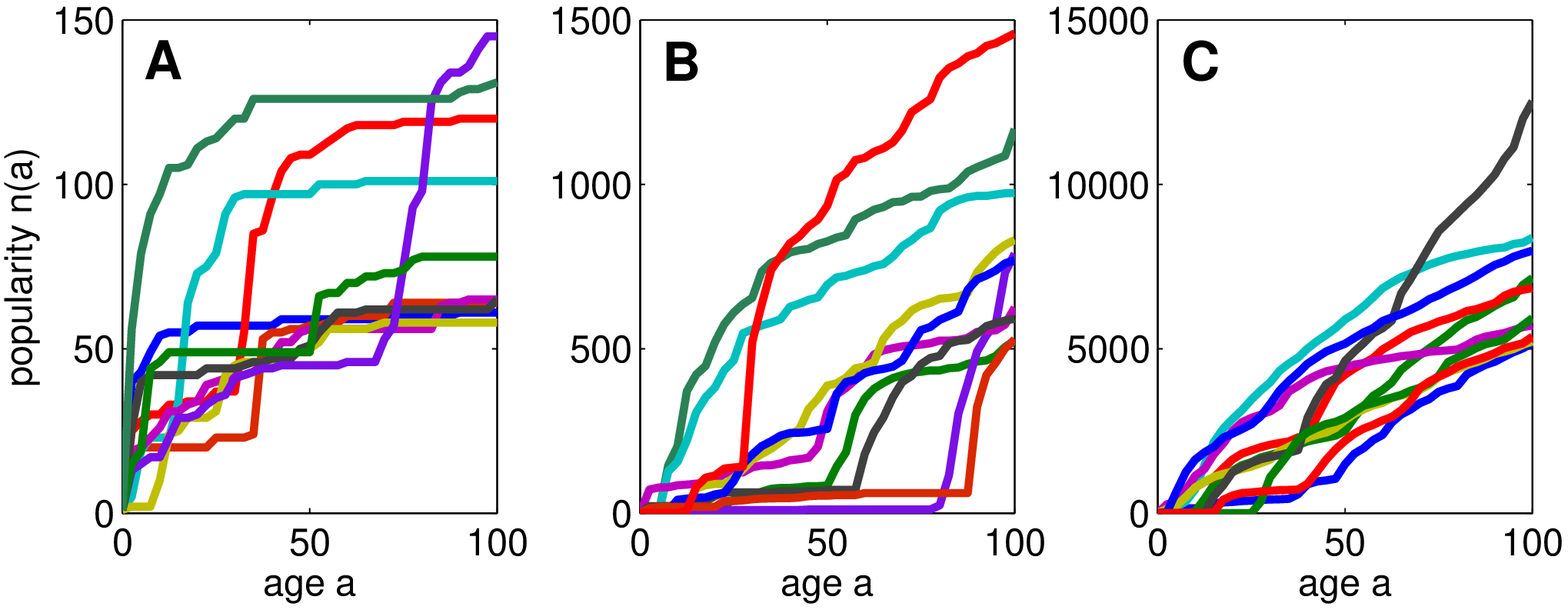,width=16.0cm}
\caption{ { Examples of the age-dependence of meme popularity from numerical simulations of the model. Each panel shows the popularity of 10 different memes; the memes plotted are chosen at random from those whose popularity at age $100$ is of order (A) $10^2$, (B) $10^3$, or (C) $10^4$. For model parameters, see the caption of Fig.~\ref{fig4}A. }}\label{fig_histories}
\end{figure}

When a user decides, at time $t$, to send a tweet, she has two options (see Fig.~1): with probability $\mu$, the user innovates, i.e., invents a new meme, and tweets this new meme to all her followers. The new meme appears in the user's own stream (it is automatically interesting to the originating user), and in the streams of all her followers (where it may be deemed interesting by each follower, independently, with probability $\lambda$). If not innovating (with probability $1-\mu$), the user instead chooses a meme from her stream to retweet. The meme for retweeting is chosen by looking backwards in time an amount $t_m$ determined by a draw from the memory-time distribution $\Phi(t_m)$, and finding the first interesting meme in her stream that arrived prior to the time $t-t_m$. The retweeted meme then appears in the streams of the user's followers (time-stamped as time $t$), but because it is a retweet, it does not appear a second time in the stream of the tweeting user. The popularity $n(a)$ of a meme is the total number of times it has been tweeted  or retweeted by age $a$, i.e., by a time $a$ after its first appearance (when it was tweeted as an innovation) \cite{Simkin07}.
{ Figure~\ref{fig_histories} shows some examples of evolving meme popularities: each panel displays the popularity $n(a)$ of a single meme as a function of its age $a$.}

The model as described is a ``neutral model'' \cite{Pinto11,Kimurabook} in the sense that all memes have the same ``fitness'' \cite{Bianconi01}: no meme has an inherent  advantage in terms of its attractiveness to users. Nevertheless, the competition between memes for the limited resource of user attention causes initial random fluctuations in popularities of memes to be amplified, and leads to the variability across memes seen in Fig.~\ref{fig_histories} and to popularity distributions with very heavy tails \cite{Bentley04}: heavier, for example, than can be generated by  models of preferential attachment or cumulative advantage type \cite{Simkin11,Simon55,Price76,Barabasi99,Newman05}. This ``competition-induced criticality'' was studied for a zero-memory ($\Phi(t_m)=\delta(t_m)$) version of this model in Ref.~\cite{GleesonPRL14}. Indeed, the results of Ref.~\cite{GleesonPRL14} can be obtained as a special case of the model described here, by setting $\Phi(t_m)=\delta(t_m)$, $\lambda=1$, $\beta_{j k}\equiv 1$, and $p_{j k}=\delta_{j z} p_k$; numerical simulation results for a closely related model were first reported in Ref.~\cite{Weng12}.

A branching process approximation \cite{Harrisbook,Iribarren11} for the model enables us to understand how the network structure (via the out-degree distribution $p_k$) and the users' memory-time distribution ($\Phi(t_m)$) affect the popularity distribution of memes. Defining $q_n (a)$ as the probability that a meme has popularity (total number of (re)tweets) $n$ at age $a$, the branching process provides analytical expressions that determine the probability generating function (PGF) \cite{Wilfbook,Newman01} of the popularity distribution,
\begin{equation}
H(a;x)=\sum_{n=1}^\infty q_n (a) x^n.\label{defH}
 \end{equation}
 The details of the derivation and analysis of the branching-process approximation are given in Sec.~\ref{derivation} and \ref{analysis}. The reader who is mainly interested in the applications of the model may jump straight to Sec.~\ref{resultssynthetic}, while noting that the most important outcome of the analysis is that in the small-innovation limit $\mu \to 0$, the model describes a critical branching process, with power-law distributions of popularity (avalanche size) \cite{Goh03,SchwartzCohen,Adami02,Zapperi95}.

\section{Derivation of branching process approximation} \label{derivation}

\subsection{Derivation of governing equations}

\begin{figure}
\centering
\epsfig{figure=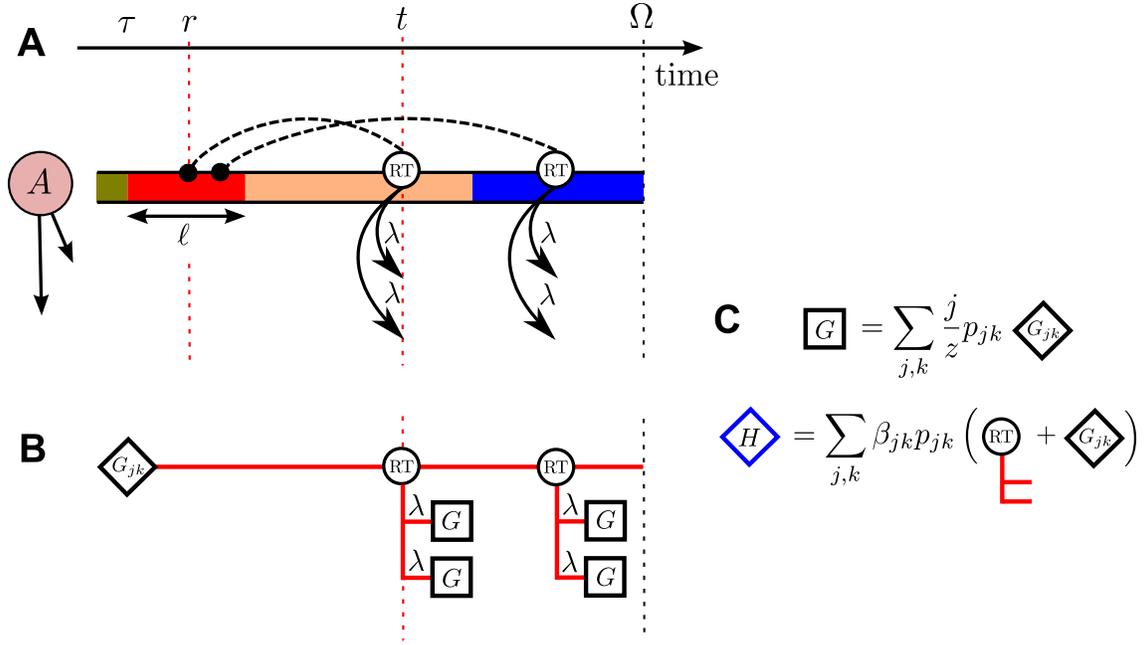,width=15.0cm}
\caption{Schematic for the derivation of the PGF equations, see Sec.~\ref{derivation}. ({A}) The stream of user $A$, showing only memes that were deemed interesting by user $A$; each color represents a different meme. At time $t$, user $A$ decides to retweet a meme from the past, and looks back to time $r$, where she finds meme $M$ (colored red). She sends this meme to her followers (not shown); each follower independently deems the meme interesting with probability $\lambda$. Also shown is a later retweet event, which also copies meme $M$. ({B}) The retweet tree for meme $M$, seeded at time $\tau$. Each retweet by user $A$ of meme $M$ generates a new branch on this tree; each branch can also generate further retweets by followers of $A$, these subtrees are denoted by squares. ({C}) Schematic depiction of Eqs.~(\ref{Gavg}) and (\ref{Hgen}).}
\label{figS1}
\end{figure}

We define $G_{jk}(\tau,\Omega;x)$ as the probability generating function for the size of the ``retweet tree'', as observed at time $\Omega$, that grows from the retweeting of a meme that entered, at time $\tau\le\Omega$, the stream of a $(j,k)$-class user, see Fig.~\ref{figS1}B. To obtain an equation for $G_{jk}$, we consider the stream of a random $(j,k)$-class user (called ``user $A$'') with a meme $M$ that entered the stream at time $\tau$ (either by innovation, or because it was received from a followed user and deemed interesting by $A$), see Fig.~\ref{figS1}A.

The likelihood that meme $M$ is retweeted in the future depends on how quickly other tweets enter the stream of user $A$. In fact, meme $M$ can be considered to ``occupy'' the stream for a time interval $\ell$ stretching from $\tau$ until the time $\tau+\ell$ when the next interesting meme enters the stream of user $A$. New memes enter the stream as a Poisson process at the constant rate\footnote{User $A$ follows $j$ users, each of which is assumed to tweet at the average rate $\overline{\beta}=\sum_{j k} \frac{k}{z} \beta_{j k}p_{j k}$.  Each meme sent by these $j$ users is deemed interesting by $A$ with probability $\lambda$, so the rate at which interesting memes enter the stream of user $A$ is $j \overline \beta \lambda$. Moreover, user $A$ innovates at a rate $\mu \beta_{j k}$, which gives the second term of Eq.~(\ref{rate}). If either an incoming tweet or an innovation event occurs, a new meme is inserted into the stream of user $A$, and the occupation time of meme $M$ is ended.\label{footnote2}}
\begin{equation}
r_{j k} = j \overline{\beta} \lambda + \mu \beta_{j k},\label{rate}
\end{equation}
so the occupation time $\ell$ of meme $M$---the time it occupies the stream of user $A$---is an exponentially distributed random variable with density
\begin{equation}
P_\text{occ}(\ell)= r_{j k} \exp\left(-r_{j k} \ell\right).\label{Pocc}
\end{equation}
We note in passing that the mean occupation time
\begin{equation}
\left< \ell \right>=\int_0^\infty \ell \,P_\text{occ}(\ell) \,d\ell=\frac{1}{j \overline{\beta} \lambda + \mu \beta_{j k}} \label{lmean}
\end{equation}
is, for small innovation probabilities $\mu$, inversely proportional to $j$, the number of users followed. Thus, a user who follows many others experiences tweets entering his stream at a higher rate than a lower-$j$ user (compare the streams of users $B$ and $C$ in the schematic Fig.~1).  Consequently, the high-$j$ user is less likely to see (and so to retweet) a given meme than a low-$j$ user. This aspect of the model  clearly reflects  empirical data, as seen in Fig.~3 of \cite{Hodas12} for example.

To determine the size of trees originating from meme $M$, we consider that trees observed at a time $\Omega\ge \tau$ must be created by the retweeting by user $A$, at some  time(s) between $\tau$ and $\Omega$, via looking back in her stream to a time $r$, where $r$ lies between $\tau$ and $\min(\tau+\ell,\Omega)$ (i.e., $r$ lies within the time interval where meme $M$ occupies the stream). Let's consider a time interval of (small) length $dr$, centered at time $r$, and calculate the size of trees that are seeded by a retweet based on a lookback into this interval, from a time $t$, with $t>r$, see Fig.~\ref{figS1}. In each $dt$ interval centered at time $t$, a tree will be seeded with probability\footnote{The factor $(1-\mu)\beta_{j k}\, dt$ is the probability that a $(j,k)$-class user becomes active in the $dt$ interval and copies rather than innovates; the factor $\Phi(t-r)\,dr$ is the probability that this user chooses to copy from the $dr$-interval. }
\begin{equation}
P_\text{seed} = (1-\mu)\beta_{j k} \Phi(t-r) \, dr\, dt,
\end{equation}
and will grow to a tree with size distribution (at observation time $\Omega$) generated by\footnote{There are $k$ followers of user $A$, each of whom may deem the tweet ``uninteresting'' with probability $1-\lambda$, or consider it ``interesting''---and accept  it into their stream---with probability $\lambda$. The factor of $x$ counts the increase in popularity due to the tweet event.}
\begin{equation}
R_{k}(t,\Omega;x)= x \left[ 1-\lambda+\lambda G(t,\Omega;x)\right]^{k},
\end{equation}
where
\begin{equation}
G(t,\Omega;x) = \sum_{j,k}\frac{j}{z}p_{jk}G_{jk}(t,\Omega;x) \label{Gavg}
\end{equation}
is the PGF for the sizes of trees originating from the successful insertion at time $t$ of a meme (that is deemed interesting) into the stream of a random follower.

To calculate the total size of the tree seeded by copying from the $dr$-interval, we must add the sizes of trees that are copied into all times $t$ with $t>r$. Since each copying event is independent, the total tree size is generated by
\begin{equation}
J(r;x) = \prod_{t=r}^\Omega \left[ 1-P_\text{seed} + P_\text{seed} R_{k}(t,\Omega;x)\right].
\end{equation}
Taking logarithms of both sides of this equation and expanding to first order in $dt$ gives
\begin{align}
\log J \
& = \sum_{t=r}^\Omega \log\left[1-(1-\mu)\beta_{j k}\Phi(t-r)\,dr \,dt (1- R_{ k}(t,\Omega;x))\right]\nonumber\\
& \approx -(1-\mu)\beta_{j k} \sum_{t=r}^\Omega \Phi(t-r) \,dr\, dt (1-R_{k}(t,\Omega;x)) \nonumber\\
&\rightarrow - (1-\mu)\beta_{j k} \, dr \int_{r}^\Omega \Phi(t-r) (1-R_{k}(t,\Omega;x))\, dt  \quad\text{ as }dt\to 0,
\end{align}
so $J(r;x)$ can be written as
\begin{equation}
J(r;x) = \exp\left[-(1-\mu)\beta_{j k}\, dr\int_r^\Omega\Phi(t-r) (1-R_{ k}(t,\Omega;x))\, dt\right]. \label{J8}
\end{equation}
Recall that $J(r;x)$ is the PGF for trees seeded by copying from time $r$. To obtain the total size of all children trees of meme $M$, we must consider trees seeded at all possible times $r$ from $\tau$ to the  time $\min(\tau+\ell,\Omega)$ that marks the end of the occupation of user $A$'s stream by meme $M$. Each $dr$ time interval again independently generates trees with sizes distributed according to Eq.~(\ref{J8}), so the PGF for the total size is found by multiplying together copies of the $J(r;x)$ function for each $dr$ time interval, thus:
\begin{align}
P_\text{size}(\ell) & = \prod_{r=\tau}^{\min(\tau+\ell,\Omega)}  J(r;x)\nonumber\\
&  =\exp\left[-(1-\mu)\beta_{j k}\sum_{r=\t}^{\min(\tau+\ell,\Omega)} dr\, \int_r^\Omega\Phi(t-r) (1- R_{k}(t,\Omega;x))\, dt\right]\nonumber\\
& \rightarrow \exp\left[-(1-\mu)\beta_{j k}\int_{\t}^{\min(\tau+\ell,\Omega)} dr\, \int_r^\Omega dt\,\Phi(t-r) (1- R_{k}(t,\Omega;x))\right]\quad \text{ as } dr\to 0.\label{P9}
\end{align}
Combining probabilities, by integrating over all possible occupation times $\ell$, gives
\begin{equation}
G_{jk}(\tau,\Omega;x) = \int_0^\infty P_\text{occ}(\ell) P_\text{size}(\ell)\, d\ell
\end{equation}
and combining Eqs.~(\ref{Pocc}), (\ref{Gavg}) and (\ref{P9}) yields an integral equation for $G$:
\begin{align}
G(\t,\Omega;x)&=\sum_{j k}\frac{j}{z}p_{j k} \int_0^\infty d\ell\, \left(j \overline{\beta}\lambda + \mu \beta_{j k}\right)\exp\left[-\left(j \overline{\beta}\lambda + \mu \beta_{j k}\right)\ell\right] \times\nonumber\\
 &\times\exp\left[-(1-\mu)\beta_{j k}\int_{0}^{\min(\t+\ell,\Omega)} dr\, \int_r^\Omega dt\,\Phi(t-r) (1- x\left[1-\lambda+\lambda G(t,\Omega;x)\right]^k)\right]. \label{E10}
\end{align}
{
Introducing the change of variables $a=\Omega-\tau$, $\tilde r = r-\tau$, $\tilde \tau = \Omega-t$, we rewrite this equation as
\begin{align}
G(\Omega-a,\Omega;x)&=\sum_{j k}\frac{j}{z}p_{j k} \int_0^\infty d\ell\, \left(j \overline{\beta}\lambda + \mu \beta_{j k}\right)\exp\left[-\left(j \overline{\beta}\lambda + \mu \beta_{j k}\right)\ell\right] \times\nonumber\\
 &\hspace{-1.0cm}\times\exp\left[-(1-\mu)\beta_{j k}\int_{0}^{\min(\ell,a)} d\tilde{r}\, \int_{0}^{a-\tilde{r}} d\tilde{\t}\,\Phi(a-\tilde{r}-\tilde{\t}) (1- x\left[1-\lambda+\lambda G(\Omega-\tilde{\t},\Omega;x)\right]^k)\right].\label{e19A}
\end{align}
Note that the only appearance of the
 observation time $\Omega$ in this equation is in the first two arguments of the $G$ function: this reflects the fact that the popularity of memes in this model depends only on their age $a$ (unlike cumulative-advantage models, which exhibit a dependence also on the global time because early-born items have an ``early-mover'' advantage \cite{Newman09}). We therefore compress the notation by defining $G$ in terms only of the age $a$ of the memes: $ G(\Omega-\tau;x):=G(\tau,\Omega;x)$, and $G(a;x)$ solves the integral equation
\begin{align}
G(a;x)&=\sum_{j k}\frac{j}{z}p_{j k} \int_0^\infty d\ell\, \left(j \overline{\beta}\lambda + \mu \beta_{j k}\right)\exp\left[-\left(j \overline{\beta}\lambda + \mu \beta_{j k}\right)\ell\right] \times\nonumber\\
 &\times\exp\left[-(1-\mu)\beta_{j k}\int_{0}^{\min(\ell,a)} d\tilde{r}\, \int_{0}^{a-\tilde{r}} d\tilde{\t}\,\Phi(a-\tilde{r}-\tilde{\t}) (1- x\left[1-\lambda+\lambda G(\tilde{\t};x)\right]^k)\right],\label{e19}
\end{align}
with initial condition $G(0;x)=1$.
}

The popularity of a meme, as observed at time $\Omega$, that is seeded by a single tweet (e.g., by an innovation) at time $\tau$ may be calculated in a similar way to the derivation of Eq.~(\ref{e19}); the generating function is of the form
\begin{equation}
H(\tau,\Omega;x) = \sum_{j,k} \beta_{j k} p_{j k} R_k(\tau,\Omega;x) G_{j k}(\tau,\Omega;x),\label{Hgen}
\end{equation}
where $\beta_{jk}p_{jk}$ represents the probability that the seed tweet originates from a $(j,k)$-class user, $R_k$ is the PGF for the trees generated from the followers of the user, and $G_{jk}$ is the PGF for the size of the retweet-tree of the meme (see Fig.~\ref{figS1}C). Introducing the age $a$ of the meme as before and defining $q_n(a)$ as the probability that an age-$a$ meme has popularity $n$, we have the PGF defined in Eq.~(\ref{defH}),
which is given by
\begin{align}
H(a;x)&=\sum_{j k} \beta_{j k} p_{j k} x \left[1-\lambda+\lambda G(a;x)\right]^k  \int_0^\infty d\ell\, \left(j \overline{\beta}\lambda + \mu \beta_{j k}\right)\exp\left[-\left(j \overline{\beta}\lambda + \mu \beta_{j k}\right)\ell\right] \times\nonumber\\
 &\times\exp\left[-(1-\mu)\beta_{j k}\int_{0}^{\min(\ell,a)} d\tilde{r}\, \int_{0}^{a-\tilde{r}} d\tilde{\t}\,\Phi(a-\tilde{r}-\tilde{\t}) (1- x\left[1-\lambda+\lambda G(\tilde{\t};x)\right]^k)\right];\label{e21}
\end{align}
the initial condition is $H(0;x)=x$ (i.e., all memes have initial popularity 1: $q_n(0)=\delta_{n,1}$).

\subsection{Distribution of response times} \label{sec3.2}
It is worth noting that all agents in the model have constant activity rates, so that the actions of each individual agent constitute a Poisson process. A Poisson process is characterized by an exponential distribution of inter-event times, where each event corresponds to an innovation or a retweeting action. This assumption is contrary to studies such as \cite{Vazquez07,Min11,Iribarren09,Iribarren11,JoPRX14,TemporalNetworks,Barabasi05,Malmgren08,Hoffmann12,Boguna14,VanMieghem13,Kivela14}, where heavy-tailed distributions of inter-event times are examined. Despite this, in our model the memory-time distribution $\Phi(t_m)$ directly influences the waiting times (or ``response times'') between the receipt of a specific meme, and the retweeting of it. Indeed, if $\Phi(t_m)$ is a heavy-tailed distribution, then a meme received by a given user at time $\tau$ will be retweeted by that user at a time $t$ (with $t\gg\tau$) with probability proportional to $\Phi(t-\tau)$ (the exact relation depends on how long the meme occupies the stream of the user).
Therefore, a heavy-tailed memory distribution  gives rise to a heavy-tailed waiting-time distribution for individual memes, despite the fact that the activity of each individual user is described by a Poisson process (cf.~the heavy-tailed waiting-time distributions found in empirical studies of email correspondence \cite{Barabasi05,Malmgren08}). It is clearly important to distinguish between the distributions of inter-event times (for actions of users) and of the waiting times experienced by individual memes: the model assumes each user has exponentially-distributed inter-event times, but it can nevertheless produce heavy-tailed distributions of waiting times for memes to be retweeted.

In particular, if the memory-time distribution $\Phi(t_m)$ is a $\text{Gamma}(k_G,\theta)$ distribution \cite{Iribarren11} as used in Secs.~\ref{resultssynthetic} and \ref{resultsdata}, i.e.,  $\Phi(t_m)=\frac{1}{\Gamma(k_G)\theta^{k_G}}t_m^{k_G-1}\exp\left(-t_m/\theta\right)$, then $\Phi(t_m)$ is approximately power-law  for memory times $t_m$ with $t_m\ll \theta$, with an exponential cutoff at larger times. The corresponding waiting-time distribution shows a similar scaling in this range, like the slow decay noted in empirical response times for Twitter users (e.g., in Fig.~5 of \cite{Hodas12}). { In Sec.~\ref{Discussion} we consider how the model could be extended to incorporate bursty (non-Poisson) user activity.}

\section{Analysis} \label{analysis}

\subsection{Criticality of the branching process} \label{criticality}
A branching process may be classified by the expected (mean) number $\xi$ of ``children'' of each ``parent'': if this number (called the ``branching number'') is less than 1, the process is \emph{subcritical} and if $\xi$ is greater than 1 the process is \emph{supercritical}. \emph{Critical} branching processes, with an average of exactly one child per parent, give rise to power-law distributions of tree-sizes and of durations of growth cascades, and have been used to examine self-organized criticality in sandpile models on networks  \cite{Goh03,Zapperi95}. Here we demonstrate that the general process derived in Sec.~\ref{derivation} is a critical branching process in the limit of vanishing innovation $\mu\to0$.

We identify the ``parent'' in the process as a meme that was accepted into the stream (i.e., deemed interesting) of a $(j,k)$-class user at time $\tau$: see, for example, meme $M$ in the stream of user $A$, as shown in Fig.~\ref{figS1}. The ``children'' of this meme are the retweets of it that are accepted into the streams of the followers of $A$ at any time $t>\tau$. The PGF for the number of children of meme $M$ is derived by following the same steps as in Sec.~\ref{derivation}, but replacing $R_k$ by $(1-\lambda+\lambda x)^k$: each power of $x$ then counts a successful insertion of meme $M$ into the stream of one of the $k$ followers of $A$. The resulting PGF, for a meme of age $a$, is (cf. Eq.~(\ref{E10}))
\begin{align}
K_{jk}(a;x)&= \int_0^\infty d\ell\, P_\text{occ}(\ell) \times\nonumber\\
 &\times\exp\left[-(1-\mu)\beta_{j k}\int_{0}^{\min(\ell,a)} d\tilde{r}\, \int_{0}^{a-\tilde{r}} d\tilde{\t}\,\Phi(a-\tilde{r}-\tilde{\t}) (1- \left[1-\lambda+\lambda x\right]^k)\right] \nonumber\\
 &= \int_0^\infty d\ell\, P_\text{occ}(\ell)
\exp\left[-(1-\mu)\beta_{j k} (1- \left[1-\lambda+\lambda x\right]^k) \int_{0}^{\min(\ell,a)} C(a-\tilde r) d\tilde{r}\right],
  \label{EK}
\end{align}
where $C(t)=\int_0^t \Phi(t_m)dt_m$ is the cumulative distribution function (CDF) for memory times. The expected (mean) number of children for a meme in the $(j,k)$-class stream is determined from the PGF in the usual way \cite{Wilfbook}, by differentiating with respect to $x$ and evaluating at $x=1$, thus:
\begin{equation}
\xi_{j k} = \left. \frac{\partial K_{j k}}{\partial x}\right|_{x=1}.
\end{equation}
In the limit of large ages, $a\to\infty$, we use the fact that $C(\infty)=1$ to obtain
\begin{align}
\xi_{jk}&\sim (1-\mu)\beta_{jk}\lambda k \int_0^\infty \ell\, P_\text{occ}(\ell) \,d\ell\quad \text{ as }a\to \infty \nonumber\\
   &= \frac{(1-\mu) \beta_{j k}\lambda k}{j \overline{\beta}  \lambda+ \mu \beta_{jk}}.\label{xijk}
\end{align}
Averaging over all $(j,k)$ classes, the effective branching number $\xi$ of the process is the expected number of children of a meme that is accepted into the stream of a random follower:
\begin{align}
\xi & = \sum_{j,k} \frac{j}{z} p_{j k} \xi_{j k} \nonumber\\
   & \to \sum_{j,k} \frac{j}{z} p_{j k} \frac{\beta_{jk}\lambda k}{j \overline \beta \lambda}  = 1 \quad\text{ as }\mu \to 0 \label{xi}
\end{align}
(recall that $\overline \beta \equiv \sum_{j,k} \frac{k}{z}\beta_{j k} p_{j k}$).

Thus, we have shown that the branching process underlying the model is critical when $\mu=0$. The  occupation time of a meme in a users' stream is due to the competition between neutral-fitness memes for the limited resource of user attention; this competition ensures that the mean number of successful retweets (children) generated during the finite  occupation time of the meme  is precisely one, and so induces the power-law distributions of cascade sizes that are characteristic of critical branching processes \cite{Goh03,Zapperi95}.

{ It is worth noting that the result of Eq.~(\ref{xi}) can also be derived in a more heuristic fashion, which enables us to discuss possible generalizations of the model in Sec.~\ref{limitations}. As above, we want to calculate $\xi_{j k}$, the expected number of children of a parent meme $M$ that has been accepted into the stream of a $(j,k)$-class user, called user $A$. We consider a (long) time window of duration $W$ units. During this time window, a total of approximately $(j \overline{\beta} \lambda + \mu \beta_{j k})W$ tweets have been accepted into the stream of user $A$ (see footnote~\ref{footnote2} and Eq.~(\ref{lmean})). When user $A$ decides to retweet during the time window, one of these memes is chosen for retweeting. If the times chosen by the user are uniformly distributed over the window then the probability that the chosen meme is meme $M$ is
\begin{equation}
P_\text{chosen} = \frac{1}{\text{number of memes in stream}} = \frac{1}{(j \overline{\beta} \lambda + \mu \beta_{j k})W} \label{Pchosen}.
\end{equation}
Alternatively, this result can be calculated by noting that the average time that a single meme occupies the stream is given by $\left<\ell\right>$ in Eq.~(\ref{lmean}), so the expected fraction of the total time that meme $M$ occupies the stream of user $A$ over the window of length $W$ is $\left<\ell\right>/W = P_\text{chosen}$.

Recalling that the activity rate of user $A$ is $\beta_{j k}$, the expected number of retweets by this user during the time window is
\begin{equation}
N_\text{retweets} = (1-\mu) \beta_{j k} W. \label{Nretweets}
\end{equation}
Each retweet is broadcast to the $k$
followers of $A$, each of whom finds the retweet interesting with probability $\lambda$, so the expected number of children (memes deemed interesting by followers) per retweet is $\lambda k$. The expected number of children of the parent meme $M$ over the time window is therefore
\begin{align}
\xi_{j k} & = \left(\text{number of retweets by $A$}\right)\times\left(\text{probability meme $M$ is chosen}\right)\times\left(\text{children per retweet}\right) \nonumber\\
& = N_\text{retweets} P_\text{chosen} \lambda k ,\label{xijkB}
\end{align}
which recovers Eq.~(\ref{xijk}). The expected number $\xi$ of children of a meme that is accepted into the stream of a random follower is then calculated as in Eq.~(\ref{xi}), giving $\xi\to 1$ in the $\mu \to 0$ limit.
}

\subsection{An explicit expression for $q_1(a)$}\label{sec:q1}
The value $q_1(a)$ is the probability that a meme, once created via an innovation event, is not retweeted by the time it reaches age $a$: recall that the popularity $n$ of a meme is set to 1 when it is first tweeted (i.e., at birth); subsequent retweets (if any) increase the value of $n$ above 1. The probability $q_1(a)$ may be calculated explicitly using Eq.~(\ref{e21}):
\begin{align}
q_1(a) &= \lim_{x\to0}\frac{H(a;x)}{x},\nonumber\\
& = \sum_{j,k} \beta_{jk}p_{jk}\left[1-\lambda+\lambda G(a;0)\right]^k  \int_0^\infty d\ell\, P_\text{occ}(\ell)
\exp\left[-(1-\mu)\beta_{j k}  \int_{0}^{\min(\ell,a)} C(a-\tilde r) d\tilde{r}\right],
\end{align}
with $G(a;0)$ given, from Eq.~(\ref{e19}), by
\begin{align}
G(a;0) &= \sum_{j,k}\frac{j}{z} p_{j k }\int_0^\infty d\ell\, P_\text{occ}(\ell)
\exp\left[-(1-\mu)\beta_{j k}  \int_{0}^{\min(\ell,a)} C(a-\tilde r) d\tilde{r}\right].
\end{align}
If we consider the large-age limit, $a\to\infty$, than we can approximate the integral of the cumulative distribution function for memory times as
\begin{equation}
\int_{0}^{\min(\ell,a)} C(a-\tilde r) d\tilde{r} \approx \ell \,C(a)
\end{equation}
and the integral over $\ell$ can be calculated to give the large-$a$ approximation
\begin{equation}
q_1(a) \sim \sum_{j k}\beta_{j k} p_{j k} \frac{j \overline{\beta} \lambda + \mu \beta_{j k}}{ j \overline{\beta}\lambda + \mu\beta_{jk} + (1-\mu) \beta_{j k}C(a)}\left[1-\lambda+\lambda G(a;0)\right]^k,\label{q1a}
\end{equation}
with
\begin{equation}
G(a;0)\sim \sum_{ j k}\frac{j}{z} p_{j k} \frac{j \overline{\beta} \lambda + \mu \beta_{j k}}{ j \overline{\beta}\lambda +  \mu\beta_{jk} + (1-\mu) \beta_{j k}C(a)}.\label{q1b}
\end{equation}
 In the simplified case  $p_{jk}=\delta_{j,z}p_k$ and $\beta_{j k}\equiv 1$, Eqs.~(\ref{q1a}) and (\ref{q1b}) reduce to
 \begin{equation}
q_1(a)\sim \frac{\lambda z+\mu}{\lambda z + \mu+(1-\mu)C(a)}\sum_{k=0}^\infty p_k \left[ 1-\lambda +\lambda \frac{\lambda z+\mu}{\lambda z+\mu+(1-\mu)C(a)}\right]^k. \label{q1}
\end{equation}

The $a=\infty$ limit of $q_1(a)$ gives the fraction of memes that are \emph{never} retweeted, and so have popularity $n=1$ forever. The value of $q_1(\infty)$ is obtained from Eqs.~(\ref{q1a}) and (\ref{q1b}) by setting $C(a)$ to its $a\to\infty$ limit of 1. The approach of $q_1(a)$ towards the value $q_1(\infty)$ depends, through the CDF $C(a)$, on the tail of the memory-time distribution $\Phi$. If the distribution $\Phi$ is heavy-tailed, there is a non-negligible probability that a meme may be retweeted even if a very long time has elapsed since its birth.

\subsection{Mean popularity} \label{sec:mean}
The age dependence of the mean popularity (i.e., the expected number of tweets/retweets for a meme of age $a$) is given by
\begin{equation}
m(a) = \sum_{n=1}^\infty n\, q_n(a) = \left.\frac{\partial H(a;x)}{\partial x}\right|_{x=1}.
\end{equation}
Differentiating (\ref{e21}) and setting $x=1$ yields an integral equation for $m(a)$:
\begin{align}
m(a) &= \sum_{j k}\beta_{j k}p_{j k}\left\{ 1+\lambda\, k\, m_G(a) + (1-\mu)\beta_{j k}\int_0^\infty
d\ell\, \left(j \overline{\beta}\lambda + \mu \beta_{j k}\right)\exp\left[-\left(j \overline{\beta}\lambda + \mu \beta_{j k}\right)\ell\right] \right.\times\nonumber\\
 &\left.\times\int_{0}^{\min(\ell,a)} d\tilde{r}\, \int_{0}^{a-\tilde{r}} d\tilde{\t}\,\Phi(a-\tilde{r}-\tilde{\t}) \left[1+\lambda \,k\, m_G(\tilde\tau)\right]\right\},\label{21}
 \end{align}
 where $m_G(a)$, defined by $m_G(a) = \left.\frac{\partial G(a;x)}{\partial x}\right|_{x=1}$, is the solution of the integral equation found by differentiating Eq.~(\ref{e19}):
 \begin{align}
 m_G(a)&=\sum_{j k}\frac{j}{z} p_{j k}\int_0^\infty
d\ell\, \left(j \overline{\beta}\lambda + \mu \beta_{j k}\right)\exp\left[-\left(j \overline{\beta}\lambda + \mu \beta_{j k}\right)\ell\right] \times\nonumber\\
 &\times(1-\mu)\beta_{j k}\int_{0}^{\min(\ell,a)} d\tilde{r}\, \int_{0}^{a-\tilde{r}} d\tilde{\t}\,\Phi(a-\tilde{r}-\tilde{\t}) \left[1+\lambda \,k\, m_G(\tilde\tau)\right].\label{22}
\end{align}
The order of the time integrals may be swapped using the identity
\begin{equation}
\int_0^\infty d\ell\int_0^{\min(\ell,a)} d\tilde r = \int_0^a d\tilde r\int_{\tilde r}^\infty d\ell,\label{swap}
\end{equation}
and the resulting $\ell$ integral can be performed explicitly:
\begin{equation}
\int_{\tilde r}^\infty (j \overline \beta \lambda+\mu \beta_{jk})e^{-(j \overline \beta \lambda+\mu \beta_{jk})\ell}d\ell = e^{-(j \overline \beta \lambda+\mu \beta_{jk})\tilde r}.
\end{equation}
As a result, the expressions (\ref{21}) and (\ref{22}) can be written as double convolution integrals. Taking Laplace transforms, Eq.~(\ref{21}) then becomes
\begin{equation}
\hat m(s) = \frac{1}{s}+z \overline \beta \lambda \hat{m}_G(s) + (1-\mu) \hat{\Phi}(s) \sum_{j,k} \beta_{j k}^2 p_{j k}\frac{\frac{1}{s}+\lambda k \hat{m}_G(s)}{j \overline \beta \lambda + \mu \beta_{j k}+s}, \label{MH}
\end{equation}
where hats denote Laplace transforms, e.g.,
\begin{equation}
\hat \Phi(s) \equiv \int_0^\infty e^{-s t} \Phi(t) dt,
\end{equation}
and with $\hat{m}_G(s)$ given explicitly from the Laplace transform of Eq.~(\ref{22}):
\begin{equation}
\hat{m}_G(s)=\frac{ (1-\mu)\hat{\Phi}(s) \sum_{j,k} \frac{j}{z}p_{jk}\frac{\beta_{jk}}{j \overline \beta \lambda + \mu \beta_{jk}+s}}{s\left[ 1-(1-\mu)\lambda \hat{\Phi}(s) \sum_{j,k} \frac{j}{z}p_{j k}\frac{k \beta_{j k}}{j \overline \beta \lambda + \mu \beta_{j k}+s}\right]}.
\end{equation}
If we specialize now to the simplified case  where $\beta_{jk}\equiv 1$ for all $(j,k)$ classes, and $p_{jk}=\delta_{j,z} p_k$, we obtain the simpler expression
\begin{equation}
\hat{m}_G(s)=\frac{ (1-\mu)\hat{\Phi}(s) \frac{1}{\lambda z +\mu+s}}{s\left[ 1-(1-\mu) \hat{\Phi}(s) \frac{\lambda z}{\lambda z +\mu+s}\right]}. \label{ma0}
\end{equation}
Substituting for $\hat{m}_G$ into the simplified version of Eq.~(\ref{MH}) yields
\begin{equation}
\hat{m}(s) = \frac{1}{s}+\frac{1-\mu}{s}\frac{(\lambda z +1)\hat\Phi(s)}{\lambda z+ \mu+s-(1-\mu)\lambda z \hat \Phi(s)}. \label{Eq1maintext}
\end{equation}
Note that, unlike the expression  for $q_1$ in Eq.~(\ref{q1}), the mean popularity depends on the out-degree distribution $p_k$ only through the mean degree $z$, implying that the mean popularity is independent of the finer details of the network structure.

To consider the large-age asymptotics of $m(a)$ from Eq.~(\ref{Eq1maintext}) we use results from renewal theory \cite{Iribarren11,AthreyaNeybook}. If the Malthusian parameter $\alpha$ exists, where $\alpha$ is defined as the solution of the equation
\begin{equation}
\frac{(1-\mu)\lambda z \hat\Phi(\alpha)}{\lambda z +\mu +\alpha}=1, \label{Malthusian}
\end{equation}
then the large-age, small-$\mu$ asymptotic behavior of $m(a)$ can be written as (Theorem IV.4.2 of \cite{AthreyaNeybook})
\begin{equation}
m(a) \sim \frac{1}{\mu}-\frac{1}{\mu} \, e^{-\frac{\mu(\lambda z+1)}{1+T \lambda z} a} \quad\text{ as } a\to \infty \text{, } \mu\to0.\label{ma2}
\end{equation}
Here we have used the fact that near criticality (i.e., as $\mu\to 0$) the Malthusian parameter $\alpha$ is determined by Eq.~(\ref{Malthusian}) to be $\alpha= -\frac{\mu(\lambda z +1)}{1+T \lambda z}+O(\mu^2)$,
where $T=\int_0^\infty t_m \Phi(t_m)\,d t_m$ is the mean memory time\footnote{Note that the Malthusian parameter exists for all the memory-time distributions considered in this paper (exponential and gamma distributions). However, if $\Phi$ is a subexponential distribution \cite{AthreyaNeybook} (such as the lognormal distribution \cite{Doerr13}), then the large-$a$ asymptotics of the mean popularity are related to the memory time CDF by
$$
m(a)\sim \frac{1}{\mu}-\frac{(1-\mu)(\lambda z+\mu)}{\mu^2 (\lambda z+1)}\left(1-C(a)\right)
$$
instead of Eq.~(\ref{ma2}).
}.
Setting $a=\infty$ in Eq.~(\ref{ma2}), we  obtain the steady-state value of the mean popularity, $m(\infty)=1/\mu$.
Although Eq.~(\ref{ma2}) is a large-$a$ asymptotic result, we may expand the exponential term about $a=0$ provided that the argument of the exponential remains small: this is valid for ages $a$ that obey the constraint
$ 
a\ll \frac{1+T\lambda z}{\mu(\lambda z +1)}.
$ 
Taking the $\mu\to 0$ limit of Eq.~(\ref{ma2}) shows that the function $m(a)$ grows linearly with $a$ for ages in this range:
\begin{equation}
m(a) \sim \frac{\lambda z +1}{1+ T \lambda z} a.\label{ma3}
\end{equation}

{
The preceding analysis all assumes that the seed node (i.e., the user who first tweets the meme of interest) is chosen at random from all the network users, with probability weighted by the user activity rate. It is straightforward to repeat the steps of the calculations for the case where the seed node is known to have $k$ followers, and so to investigate the importance of the connectivity of the seed node. Restricting our attention to the simplified case as above, and taking the infinite-age limit, we find that the expected popularity for a meme that is initiated by a seed node of out-degree $k$ is
\begin{equation} 
m_k(\infty)= \frac{\lambda z+1}{\lambda z +\mu}\left( 1 + \frac{\lambda(1-\mu)}{\mu(\lambda z+1)}k\right).
\end{equation}
Note the linear dependence of this expression on the number of followers $k$ of the seed node: memes tweeted by users with a large number of followers are likely to become more popular than memes seeded by less influential nodes. This feature of the model matches well to the observed dependence of the size of information cascades on the connectivity of the initial seed (e.g., Fig.~2 of \cite{Banos13}). Of course, the earlier results for randomly-chosen seeds are recovered by averaging over all possible seed nodes: $m(\infty)=\sum_k p_k m_k(\infty) = 1/\mu$.
}

\subsection{Infinite-age limit of popularity distribution} \label{sec:S1.6}
In the infinite-age (steady-state) limit $a \to \infty$, we assume $G(a;x) \to G_\infty(x)$, independent of $a$, and use the fact that $\int_0^\infty \Phi(t) \, dt =1$ in Eq.~(\ref{e19}) to obtain
\begin{align}
G_\infty(x)&=\sum_{j k}\frac{j}{z}p_{j k} \int_0^\infty d\ell\, \left(j \overline{\beta}\lambda + \mu \beta_{j k}\right)\exp\left[-\left(j \overline{\beta}\lambda + \mu \beta_{j k}\right)\ell\right] \times\nonumber\\
 &\times\exp\left[-(1-\mu)\beta_{j k} \ell (1- x\left[1-\lambda+\lambda G_\infty(x)\right]^k)\right].
\end{align}
Calculating the $\ell$ integral then gives the equation satisfied by $G_\infty(x)$:
\begin{equation}
G_\infty(x) = \sum_{j k}\frac{j}{z}p_{j k} \frac{ j \overline{\beta}\lambda + \mu \beta_{j k}}{ j \overline{\beta}\lambda +  \beta_{j k}-(1-\mu)\beta_{j k}  x\left[1-\lambda+\lambda G_\infty(x)\right]^k}.\label{e24}
\end{equation}
Similarly, the infinite-age limit for $H$ is given in terms of $G_\infty$ by
\begin{equation}
 H_\infty(x)= \sum_{j k}\beta_{j k} p_{j k} \frac{\left( j \overline{\beta}\lambda + \mu \beta_{j k}\right) x\left[1-\lambda+\lambda G_\infty(x)\right]^k}{ j \overline{\beta}\lambda +  \beta_{j k}-(1-\mu)\beta_{j k}  x\left[1-\lambda+\lambda G_\infty(x)\right]^k}. \label{e25}
\end{equation}
Note that these steady-state equations are independent of the memory distribution function $\Phi$. Accordingly, the asymptotic analysis approach used in \cite{GleesonPRL14} to obtain the large-$n$ behavior of the popularity distribution $q_n(\infty)$ may also be applied here: this is based on writing $x=1-w$ and $G_\infty=1-\phi(w)$ and analyzing the small-$w$, small-$\phi$ asymptotics of Eqs.~(\ref{e24}) and (\ref{e25}). We refer to \cite{GleesonPRL14} for details, and here summarize the main results for the simplified case $\beta_{j k}\equiv 1$, $p_{jk}=\delta_{j,z} p_k$.
\begin{itemize}
\item{\bf Case 1: $p_k$ has finite second moment}\\
The large-$n$ scaling of the popularity distribution is given by a power-law with exponential cutoff:
\begin{equation}
q_n(\infty)\sim A \,n^{-\frac{3}{2}} e^{-\frac{n}{\kappa}}\quad\text{ as }n\to \infty,
\end{equation}
where the prefactor $A$ is\footnote{The values of $A$, $\kappa$ and $B$ reported here are not identical to those reported in \cite{GleesonPRL14}; this is because of an approximation made in the analysis of \cite{GleesonPRL14} that is not required here (see Eq.~(S6) of \cite{GleesonPRL14}). However, the differences are of order $1/(\lambda z)$, and so are negligible in the case  $\lambda z \gg 1$ that is considered in \cite{GleesonPRL14}.}
\begin{equation}
A=\frac{z(\lambda z+1)}{\lambda z+\mu}\left[ 2\pi \left(\frac{\left< k^2 \right>(2+\lambda z-\mu)}{\lambda z +\mu}-z\right)\right]^{-\frac{1}{2}}
\end{equation}
and the cutoff $\kappa$ is
\begin{equation}
\kappa = \frac{2 \lambda^2(1-\mu)^2}{\mu^2(\lambda z+1)^2}\left[\frac{\left< k^2 \right>(2+\lambda z-\mu)}{\lambda z +\mu}-z\right].
\end{equation}
Note that $\kappa$ is proportional to $1/\mu^2$ for small $\mu$, so in the limit of vanishing innovation probability the exponential cutoff tends to infinity and the power-law part of the popularity distribution extends to all $n$.
\item{\bf Case 2: $p_k\sim D \,k^{-\gamma}$ as $k\to \infty$, with $\gamma$ between 2 and 3}\\
Immediately taking the $\mu\to 0$ limit, we find in this case that the popularity distribution has a power-law form with exponent $\gamma/(\gamma-1)$ lying between $3/2$ and $2$ \cite{Goh03,SchwartzCohen}:
\begin{equation}
q_n(\infty) \sim B \,n^{-\frac{\gamma}{\gamma-1}} \quad\text{ as }n\to \infty \label{steadystate}
\end{equation}
with prefactor $B$ given by
\begin{equation}
B=-(\lambda z+1)\frac{(D \Gamma(1-\gamma))^{-\frac{1}{\gamma-1}}}{\lambda \Gamma\left(\frac{1}{1-\gamma}\right)}\left[(\lambda z)^2\sum_{n=1}^\infty \frac{n^{\gamma-1}}{(\lambda z+1)^{n+1}}\right]^{-\frac{1}{\gamma-1}}, \label{Bprefactor}
\end{equation}
where $\Gamma$ is the gamma function.
\end{itemize}

\subsection{Large-$a$, large-$n$ asymptotics of popularity distribution}\label{sec:7.1}
{ In Appendix~\ref{AppA} we consider how the popularity distribution $q_n(a)$ behaves for large, but finite, ages, focussing on the case $\beta_{j k}\equiv 1$, $p_{j k}=\delta_{j,z} p_k$ for simplicity. The result of the asymptotic analysis is an expression for the Laplace transform of the PGF $H(a;x)$ that is valid in the $a\to\infty$ limit, see Eqs.~(\ref{H1}) and (\ref{Hasym2}) for the cases of out-degree distributions $p_k$ that have second moments $\left<k^2\right>$ that are, respectively, infinite or finite.
}

\section{Results: numerical simulation}\label{resultssynthetic}

\begin{figure}
\centering
\epsfig{figure=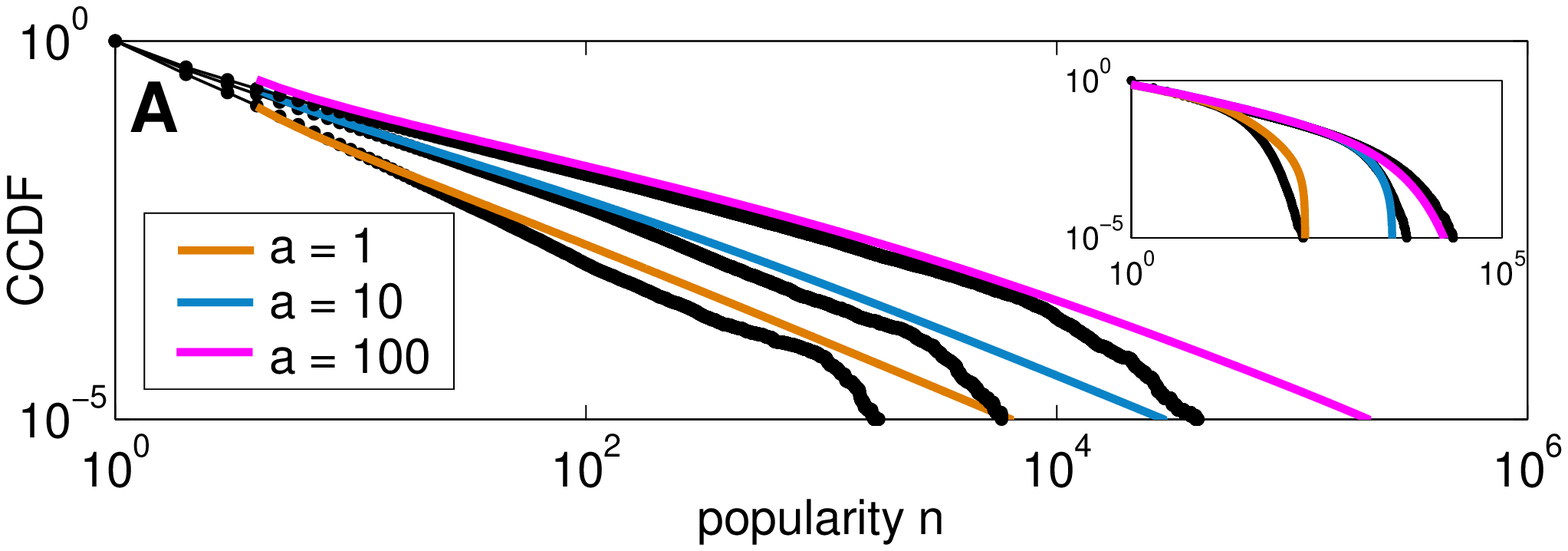,width=12.0cm}\\
\epsfig{figure=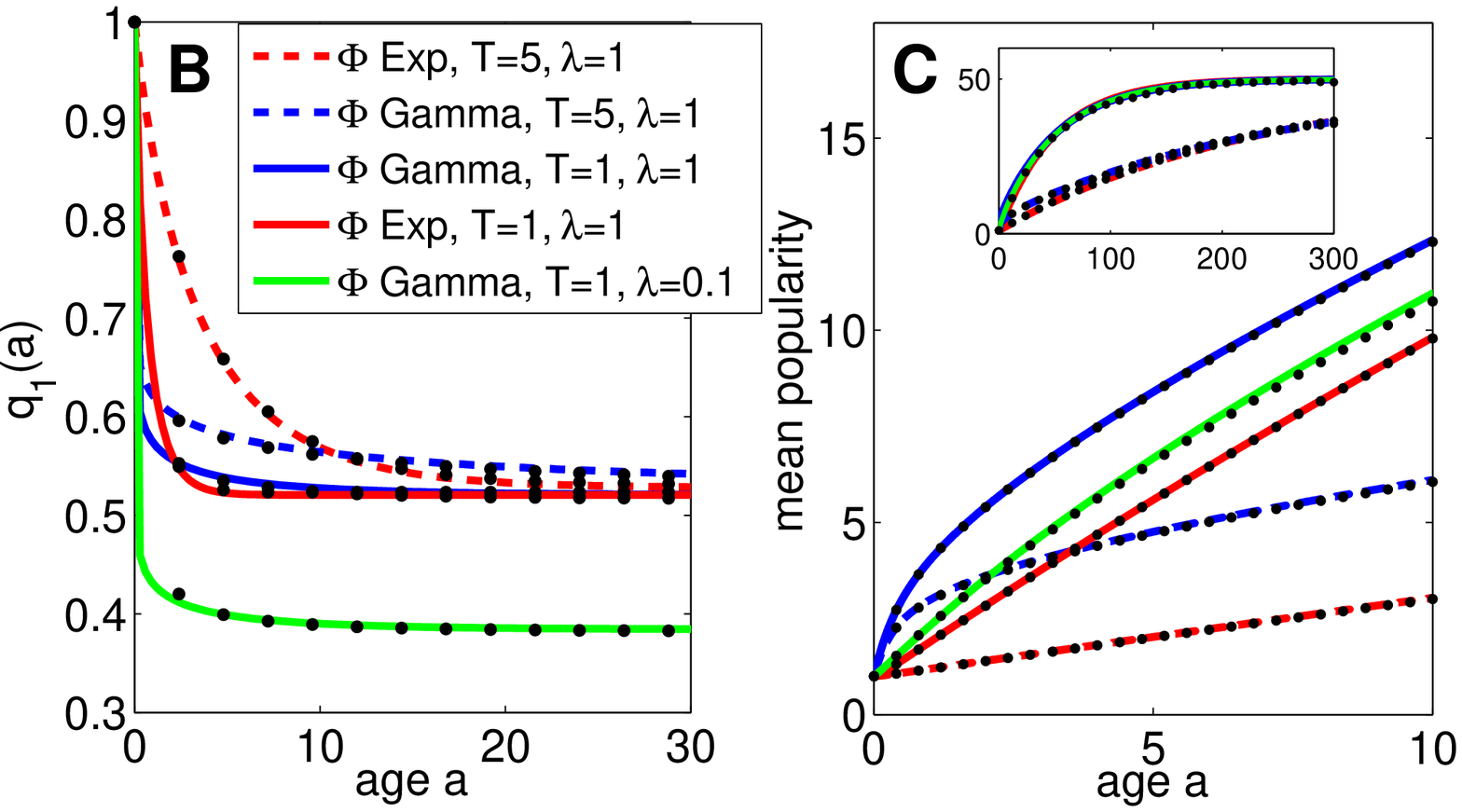,width=12.0cm}\\
\epsfig{figure=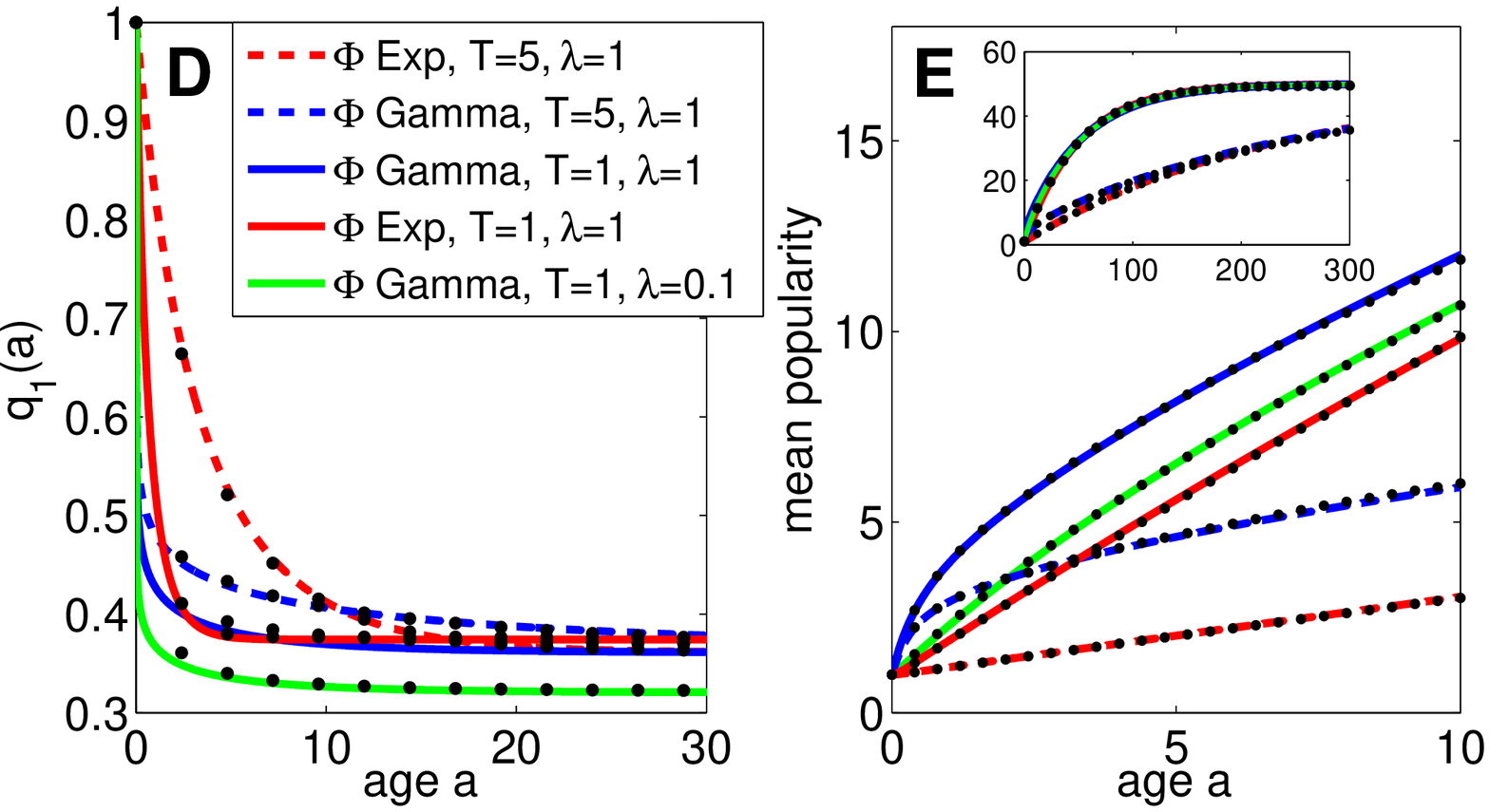,width=12.0cm}
\caption{ Numerical simulations of the model, compared with analytical results. {(A)} Complementary cumulative distribution functions (CCDFs) for meme popularity at age $a$: numerical simulation results (black) on a network with scale-free out-degree distribution ($p_k\propto k^{-\gamma}$ for $k\ge 4$  with $\gamma=2.5$, mean degree $z=11$, $N=10^5$ nodes), compared with asymptotic model result Eq.~(\ref{H1}) (colored curves). The memory-time distribution is $\Phi=\text{Gamma}(k_G,\theta)$ with $k_G=0.1$ and $\theta=50$, so the mean memory time is $T=k_G\, \theta=5$. Inset: As main, but for Poisson out-degree distribution $p_k$ ($z=11$) and gamma memory-time distribution with $k_G=0.1$ and $\theta=0.5$. {(B)} Fraction $q_1 (a)$ of memes that are not retweeted by age $a$, on the scale-free network of (A) and for various   memory-time distributions $\Phi(t_m)$ (red = exponential with mean $T$, blue/green = $\text{Gamma}(0.1,10T)$), using Eq.~(\ref{q1}). Dashed lines show the $T=5$ cases; solid lines represent $T=1$. {(C)} Mean popularity of memes of age $a$, for the same cases as in (B), and compared with Eq.~(\ref{Eq1maintext}) (using the numerical Laplace transform inversion described in Appendix~\ref{sec:inversion}); inset shows the large-$a$ behavior. All panels have $\mu=0.02$ and (except for green curves) $\lambda=1$.
(D), (E): As panels B and C , but for a network with Poisson out-degree distribution (mean degree $z=11$), with $\mu=0.02$ and (except for green curves) $\lambda=1$.}\label{fig4}
\end{figure}
To confirm the accuracy of the branching-process approximation and to explore the interactions of the network structure and the memory-time distribution, we here compare numerical simulations of the model with the theoretical predictions of Sec.~\ref{analysis}.
We generate  configuration-model directed networks with prescribed out-degree distribution $p_k$. Each one of $N$ users (nodes) is assigned a random number $k$ (drawn from the distribution $p_k$) of out-links (links to followers). The identities of the $k$ followers are chosen uniformly at random from the set of all users; in the $N\to\infty$ limit, this gives a Poisson in-degree distribution $p_j$ which, for sufficiently large $z$, gives similar results to using the in-degree distribution $p_j=\delta_{j,z}$, i.e., assuming every user follows exactly $z$ others \cite{GleesonPRL14}. Each user has the same activity rate, so $\beta_{jk}\equiv 1$.

Figure~\ref{fig4}A shows the fraction of memes that have popularity greater than or equal to $n$, at age $a$. Black symbols are the results of numerical simulations; the colored curves are determined from the large-$a$, large-$n$, $\mu=0$ asymptotics of Eq.~(\ref{H1}), using the Laplace transform inversion described in Appendix~\ref{sec:inversion}. The main figure in panel Fig.~\ref{fig4}A shows results for networks with the scale-free out-degree distribution $p_k\sim D k^{-\gamma}$ for $k\ge 4$ and exponent $\gamma=2.5$ (with $p_k=0$ for $k<4$); the inset shows the results for networks with a Poisson out-degree distribution with mean degree $z=11$ matching that of the scale-free networks. The memory time distribution is $\Phi=\text{Gamma}(k_G,\theta)$ with $k_G=0.1$, $\theta=50$ for the scale-free case and $k_G=0.1$, $\theta=5$ for the Poisson case; the mean memory time for this distribution is $T=k_G \theta$.

Panels B and C of Fig.~\ref{fig4} show results for various memory time distributions $\Phi$ on networks with the same scale-free out-degree distribution as used in panel A, and panels D and E show the corresponding results for the Poisson network. Panels B and D show the fraction $q_1(a)$ of memes that have not been retweeted by age $a$, along with the large-$a$ asymptotics of Eq.~(\ref{q1}). The age-dependence of $q_1(a)$ is qualitatively similar in panels B and D: note in both panels that the cases with longer mean memory time $T=5$ (dashed curves) approach their $a\to\infty$ limit more slowly than the $T=1$ cases (solid curves). However, the limiting value of $q_1(a)$ as $a\to\infty$ is different in the two panels, reflecting the effect of the network structure (out-degree distribution). Using Eq.~(\ref{q1}) (with $C(\infty)=1$) we obtain $q_1(\infty)=0.50$ for the scale-free network with $\lambda=1$, whereas $q_1(\infty)=0.37$ for the Poisson network.

The mean popularity $m(a)$ of age-$a$ memes is shown in panels C and E for the scale-free and Poisson networks, respectively, and for the same memory-time distributions as used in panels B and D. In contrast to the results for $q_1(a)$, we see that the finer details of the network structure have no effect on the $m(a)$ curves: panels C and E are identical, because Eq.~(\ref{Eq1maintext}) depends on $p_k$ only through the mean degree $z$, which is identical for both networks. The mean memory time $T$ determines the rate of linear growth of $m(a)$ at intermediate ages (see Eq.~(\ref{ma3})), while at early ages, the gamma memory time distribution $\Phi(t_m)$ (which has significant probability mass at low values of $t_m$) gives a faster-than-linear growth of $m(a)$ that is not present for the exponentially-distributed memory times. The large-age asymptotics  are shown in the insets; as discussed in Sec.~\ref{sec:mean}, we find $m(a)\to1/\mu$ as $a\to\infty$. { As we show in Sec.~\ref{resultsdata} below, the $m(a)$ curves can be fitted to empirical data on the popularity of Twitter hashtags; note also that the qualitative features identified here (nonlinear early growth; linear intermediate-time growth, saturation at later times) have also been observed in several other measures of information spread on social networks, such as views of YouTube videos \cite{Szabo10} and the installation of Facebook apps \cite{GleesonPNAS14}.
}

\section{Results: Twitter hashtags data}\label{resultsdata}

\subsection{Data and model inputs}
To test the ability of the model to fit real-world data, we use  a 1-year dataset comprised of the popularities of $1.4\times 10^5$ hashtags related to the 2011 15M protest movement in Spain that were tracked over the 1-year period from March 2011 to March 2012 \cite{Borge-Holthoefer11,Gonzalez-Bailon11}. We use all hashtags  for which we have at least 200 days of data; each curve in Fig.~\ref{fig6}A shows the popularity distribution for all hashtags which have the same age (to the nearest day).

\begin{figure}
\centering
\epsfig{figure=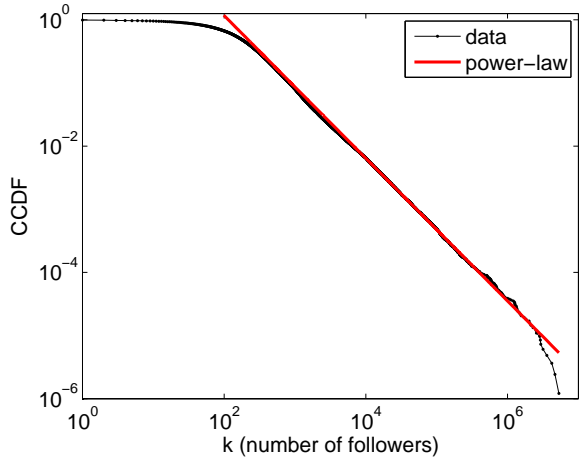,width=8.5cm} 
\caption{CCDF for the number of followers $k$ of a random sample of $8.2\times 10^5$ Twitter users. The straight line corresponds to an out-degree distribution with tail scaling as $p_k \sim D\, k^{-\gamma}$ as $k\to\infty$, with $D=240$ and $\gamma=2.13$ ($x_\text{min}=1.1\times10^4$, fitted as described in \cite{Clauset09}).}
\label{figSpk}
\end{figure}
The out-degree distribution $p_k$ of the Twitter network is an important input to the model. We determine the empirical distribution by randomly selecting $8.2\times 10^5$ Twitter user ids and recording the number of followers $k$ of each user. The measured mean number of followers is $z=703$, but the distribution $p_k$ is heavy-tailed. The complementary cumulative distribution function (CCDF) of the $k$ values is shown in Fig.~\ref{figSpk}, along with the line $D/(\gamma-1) k^{1-\gamma}$ with $D=240$ and $\gamma=2.13$ that corresponds to an out-degree distribution with tail scaling as $p_k \sim D \,k^{-\gamma}$ as $k\to\infty$ \cite{Clauset09}.

The model parameter $\lambda$ and the memory-time distribution $\Phi(t_m)$ cannot be directly estimated from the data because in cases where users receive multiple copies of the same meme (hashtag) prior to retweeting it, it is impossible to tell which of received memes ``caused'' the retweet.
Therefore, we instead use the analytical results of the model (Eqs.~(\ref{Eq1maintext}) and (\ref{H1})) to find parameter values that fit the model to the statistical characteristics of the data. Guided by the faster-than-linear growth of the mean popularity at early ages $a$ (Fig.~\ref{fig6}C) and the results of Sec.~\ref{resultssynthetic}, we assume that the memory time distribution $\Phi$ is a Gamma$(k_G,\theta)$ distribution, and fit the distribution parameters $k_G$ and $\theta$, as well as the model parameters $\mu$ and $\lambda$ to give the results in Fig.~\ref{fig6}C. Note that a delta-function memory-time distribution, as used in the toy model of \cite{GleesonPRL14}, leads to a purely-linear dependence $m(a)\propto a$, and so cannot fit to the early-time growth of the observed mean popularity.

The data does, however, provide an upper bound on the value of the innovation probability $\mu$. Recall that $\mu$ is defined as the probability that a tweeted meme (hashtag) is an innovation, i.e., that the hashtag has never before appeared in the system. Each innovation event thus increases by one the number of distinct hashtags that appear in the dataset, whereas a non-innovative (copying) tweet  will instead increase the number of copies of a hashtag that is already present in the dataset. We can therefore calculate an upper bound on the empirical innovation probability from the ratio
\begin{equation}
\tilde \mu = \frac{\text{number of distinct hashtags used in the dataset}}{\text{total number of hashtags tweeted by users}}=\frac{322799}{5886837} = 0.055. \label{mubound}
\end{equation}
Note this upper bound is consistent with the parameter value of $\mu=0.033$ that is fitted in Fig.~\ref{fig6}. The reason why Eq.~(\ref{mubound}) gives an upper bound rather than an exact value for $\mu$ is the finite size of the dataset: the data collection started at a specific point in time and so any hashtags that are in fact copied from tweets received prior to the start date will be erroneously counted as ``distinct hashtags'' in the estimate, thus leading to an overestimate of the true innovation probability.

\subsection{Results using identical user activity rates}
\begin{figure}
\centering
\epsfig{figure=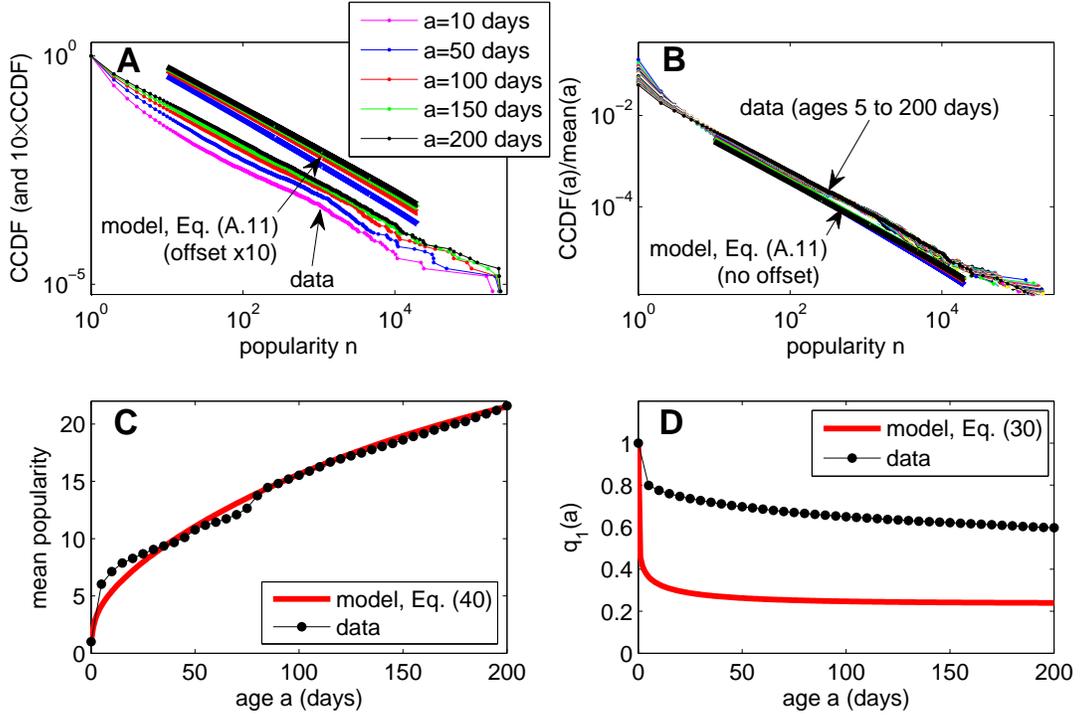,width=16.5cm}
\caption{Comparison of the model with Twitter hashtags data. {(A)} CCDFs for popularity of hashtags at age $a$ (at time $a$ after their first appearance in the dataset). The model CCDFs (from Eq.~\ref{H1}) are multiplied by 10 for clarity. Model parameters are: $\lambda=4.5\times 10^{-4}$, $\mu=0.033$, $k_G=0.25$, $\theta=500$, with one model time unit corresponding to $0.16$ days {(B)} CCDFs at age $a$, each divided by the mean popularity at age $a$. The data shows an apparent collapse onto a single curve that is closely matched by the model. {(C)} The mean popularity of hashtags of age $a$. {(D)} The fraction $q_1 (a)$ of hashtags that are not retweeted by age $a$. Here, our basic model with homogeneous user activity rates does not fit well to the data (but see Fig.~\ref{figS2}).} \label{fig6}
\end{figure}

Using the empirical Twitter out-degree distribution $p_k$, we apply the analytical results of Eqs.~(\ref{Eq1maintext}) and (\ref{H1}) (which assume $\beta_{jk}\equiv 1$) to fit the model to the data in Fig.~\ref{fig6}. Figure \ref{fig6}A and \ref{fig6}B show that the model-predicted age-dependent popularity distributions match reasonably well to the data, and Fig.~\ref{fig6}C shows that the age-dependent mean can be fitted very closely by the model. The data collapse  seen in Fig.~\ref{fig6}B is intriguing, and we analyze it further in Sec.~\ref{datacollapse} below.

Despite these successes, it was not possible to successfully fit the $q_1(a)$ curve (Fig.~\ref{fig6}D) using the simplified version of the model in which all users have the same activity rates. In Sec.~\ref{hetactivityrates} below, we therefore investigate the effect of heterogeneous activity rates, and show that an improved fit can be obtained using more realistic rates.

\subsection{Analysis of the data collapse in Fig.~\ref{fig6}B}\label{datacollapse}
As shown in Fig.~\ref{fig6}B, the ratio $q_n(a)/m(a)$ is approximately independent of the age $a$, giving a collapse of the popularity distribution data (and of the model predictions) onto a single curve.
As in Sec.~\ref{sec:7.1}, the large-$n$ asymptotics of the popularity distribution are found from the small-$w$ expansion (with $w=1-x$) of $h(a;x)=1-H(a;x)$, and  for the scale-free out-degree distribution we obtain  from Eq.~(\ref{eS56}) (using the final value theorem for Laplace transforms) the following asymptotic behavior in the  $a\to\infty$ limit:
\begin{equation}
h(\infty;1-w) \sim (\lambda z+1)C^{-\frac{1}{\gamma-1}} w^\frac{1}{\gamma-1} \quad\text{ as }w\to 0.
\end{equation}
Understanding the large-$a$ approach to this steady state (i.e., the case where $a$ is large but finite) is a difficult problem in asymptotic analysis, involving the double limits $n\to\infty$ and $a\to\infty$. However, some insight can be obtained by factoring the function $h$ into a product of its infinite-age limit $h(\infty;x)$ and another function $h_1$, with $h_1$ limiting to 1 as $a\to\infty$:
\begin{equation}
h(a;x)=h(\infty;x) h_1(a;x). \label{hh1}
\end{equation}
Taking Laplace transforms gives
\begin{equation}
\hat h(s;x)=h(\infty;x) \hat{h}_1(s;x),
\end{equation}
where
\begin{equation}
\hat{h}_1(s;1-w)= \frac{\lambda z(s+\lambda z + \hat \Phi(s))}{s(\lambda z+1)(s+\lambda z)}\frac{(\gamma-1)\lambda D^{\frac{1}{\gamma-1}}\left[\Gamma(1-\gamma)\right]^\frac{1}{\gamma-1} w^\frac{\gamma-2}{\gamma-1}\hat\Phi(s)}{s+\lambda z-\lambda z \hat\Phi(s)+(\gamma-1)\lambda D^{\frac{1}{\gamma-1}}\left[\Gamma(1-\gamma)\right]^\frac{1}{\gamma-1} w^\frac{\gamma-2}{\gamma-1}\hat\Phi(s)}.
\end{equation}
In particular, note that $\hat{h}_1(s;1-w)$ depends on $w$ only through the factor $w^\frac{\gamma-2}{\gamma-1}$. In the case where $\gamma$ is very close to 2, the exponent $(\gamma-2)/(\gamma-1)$ of the $w$ dependence is close to zero, and the dependence of $h_1$ on $w$ is therefore very weak. It follows that the rate of approach of the corresponding distribution $q_n(a)$ to the steady state $q_n(\infty)$ does not show a strong dependence on $n$, and the CCDFs for various ages appear almost parallel in the log-log plot of Fig.~\ref{fig6}A (note $\gamma=2.13$ in the Twitter network).

As we saw in Sec.~\ref{sec:mean} for the large-age asymptotics of the mean popularity, the long-time behavior of the popularity distribution may be obtained by examining the linear (early-age) growth of the  inverse transform of Eq.~(\ref{H1}). The resulting popularity distributions $q_n(a)$ show (for large $n$) a regime of linear-in-age growth, and in the case where $\gamma \approx 2$, the rate of this growth depends only weakly on $n$. Since the mean popularity $m(a)$ is also growing linearly during this age period (see Eq.~(\ref{ma3})), the division of the CCDFs at various ages by the corresponding mean $m(a)$ leads to the collapse of the data onto the single curve that is seen in Fig.~\ref{fig6}B.

\subsection{Heterogeneous activity rates} \label{hetactivityrates}
Although our analysis methods are quite general, in order to focus on understanding  the combined effects of memory and out-degree distribution most of our results thus far are specialized to the case of uniform user activity rates, $\beta_{jk}\equiv 1$. It is interesting, therefore, to examine the impact that more realistic heterogeneous activity rates would have upon the results we have obtained. To this end, we extend here to the case where the activity rate of a user depends on its out-degree $k$ while retaining the assumption $p_{jk}=\delta_{j,z}p_k$, so that $\beta_{jk} = \beta_k$ (normalized so that $\sum_k \beta_k p_k = 1$ and with $\overline \beta = \sum_k \frac{k}{z} \beta_k p_k$).

The mean popularity is given in the general case by Eq.~(\ref{MH}). Repeating the asymptotic analysis of leading to Eq.~(\ref{ma3}) for the $\mu\to0$ limit, we again find linear growth of $m(a)$ with age $a$, with a slope that generalizes that found in Eq.~(\ref{ma3}):
\begin{equation}
m(a) \sim \frac{\lambda z \overline \beta+\frac{\overline{\beta^2}}{\overline\beta}}{T \lambda z \overline \beta +1}\, a\quad\text{ as }a\to\infty,\label{ma4}
\end{equation}
where we have introduced the notation $\overline{\beta^2}\equiv \sum_k\frac{k}{z}\left( \beta_k\right)^2 p_k$.

If we additionally assume that the user activity rates saturate to a constant level $\beta_\infty$ at very large $k$, so  $\beta_k \to \beta_\infty$ as $k\to \infty$, then we can repeat the asymptotic approximations of Sec.~\ref{sec:7.1} to determine a generalized version of Eq.~(\ref{H1}):
\begin{align}
\hat H(s;x) &\sim\frac{1}{s}\nonumber\\
&\hspace{-0.3cm}- \frac{1}{s} \frac{ \lambda z \overline\beta\left(s+\lambda z \overline\beta+\frac{\overline{\beta^2}}{\overline\beta}\hat\Phi(s)\right)(\gamma-1)(1-x)\hat\Phi(s)}
{(s+\lambda z \overline\beta)\left(s+\lambda z \overline\beta-\lambda z \overline\beta \hat\Phi(s)+\beta_\infty^\frac{1}{\gamma-1}(\gamma-1)\lambda D^\frac{1}{\gamma-1}\left[\Gamma(1-\gamma)\right]^{\frac{1}{\gamma-1}}(1-x)^\frac{\gamma-2}
{\gamma-1}\hat\Phi(s)\right)}.
\end{align}

\begin{figure}
\centering
\epsfig{figure=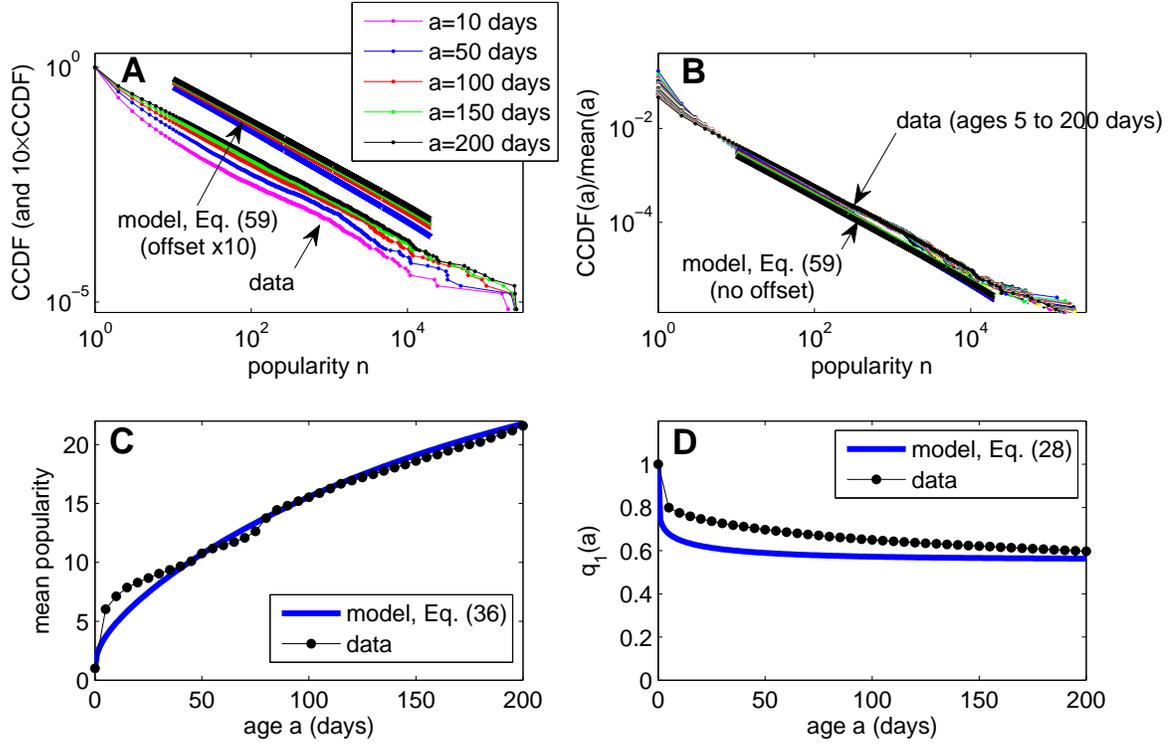,width=17cm}
\caption{As Fig.~\ref{fig6}, but including heterogeneous activity rates $\beta_k$ given by Eq.~(\ref{betak}). Model parameters are: $\lambda=5\times10^{-4}$, $\mu=0.033$, $k_G=0.25$, $\theta=500$, with one model time unit corresponding to 0.18 days. }
\label{figS2}
\end{figure}
To demonstrate the effect of heterogeneous activity rates, we consider a model for $\beta_k$ inspired by the data analysis shown in Fig.~6(a) of \cite{Hodas13}, see Appendix~\ref{AppC} for details.
Using this heterogeneous activity rate, Fig.~\ref{figS2} shows results that correspond closely to  the homogeneous-activity example of Fig.~\ref{fig6}. A comparison of panels D from both figures  clearly shows that including heterogeneous activity rates leads to a better fit of the model to the data on the fraction $q_1(a)$ of non-retweeted memes. However, the other results of the model (panels A, B and C of Fig.~\ref{figS2} compared to same panels in Fig.~\ref{fig6}) are relatively unaffected by the activity rate, so that the good matches between model and data seen in Fig.~\ref{fig6} are not compromised by including heterogeneity in activity rates.

{
\section{Limitations of the model}\label{limitations}

As we have demonstrated, the analytical tractability of the null model enables it to be fitted to time-dependent data on meme popularity. However, we were required to make a number of assumptions to obtain analytical results and in this section we briefly highlight the most important assumptions and discuss possible extensions to the model.

The network structure is assumed to be that of a directed configuration-model graph defined by the joint probability $p_{j k}$ of a node having in-degree $j$ and out-degree (number of followers) $k$. While this joint probability can encode correlations between the number followed by, and the number of followers of, a node, it does not incorporate edge-based correlations, i.e., the probability that a user with many followers is followed by users  who also have high numbers of followers. It may be possible to extend the analysis of the model to deal with at least some types of edge correlation \cite{Boguna05,Hurd16}, but this would be at the cost of increased complexity of the equations.

A more unrealistic simplification of the configuration model is the fact that it generates networks that are locally tree-like, with few short cycles. In particular, our model does not include bidirectional edges (i.e., reciprocated following relationships, where user $A$ follows user $B$ and $B$ also follows $A$), which are quite common in the Twitter network \cite{Kwak10}, but which violate the independence assumption of a branching process. However, numerical simulations in Ref.~\cite{GleesonPRL14} using a real Twitter network for a zero-memory version of the model (Sec.~S4 of \cite{GleesonPRL14}) gave quite good agreement with  branching process theory, despite the presence of a large fraction of reciprocal links in the graph. The conditions under which tree-based theories give good approximations for dynamics  on non-tree-like networks remains an active area of research \cite{Melnik11} and more work is required for further understanding.

An important assumption of the null model is that all memes have equal fitness. This is consistent with random-copying models of human decision-making \cite{Bentleybook,Bentley11} where the quality of the product---here, the ``interestingness'' of the meme---is less important than the social influence of peers' decisions \cite{Salganik06}. This neutrality of the model is at the root of the criticality of the dynamical system \cite{Pinto11}. A related (discrete-time) model for the number of citations gained by scientific papers was analyzed in Ref.~\cite{Simkin07}, where the authors also extended their neutral model to include unequal fitnesses of papers. It is likely that our model could be extended in a similar way, to incorporate a fitness parameter for each individual meme. Based on the results of Ref.~\cite{Simkin07}, we expect that our main results would be qualitatively unaffected if the distribution of fitness values over the set of all memes is strongly peaked (i.e., if most memes have roughly equal fitness values, with only the high-fitness outliers demonstrating supercritical popularity growth).

Perhaps the most unrealistic aspect of the current model is the assumption that all users have
constant activity rates, so their tweeting activity is described by a Poisson process (see the discussion in Sec.~\ref{sec3.2}). It would be interesting to relax this assumption, for example to allow the activity of users to be described by models such as that of Ref.~\cite{Perra12} or by inhomogeneous Poisson processes: the latter incorporates time-varying activity rates and so could model the 24-hour variability in tweeting levels determined by daily patterns \cite{Malmgren08}. However, we believe that the near-critical aspect of the model will not be strongly affected by such generalizations. To see this, consider the heuristic derivation of the branching number $\xi$ that was described at the end of Sec.~\ref{criticality}. Over a sufficiently long time window $W$, the expected number of interesting memes received into the stream of a $(j,k)$-class user is linear in the number $j$ of users followed, and this remains true even for inhomogeneous Poisson (or even non-Poisson) activities, provided that the observation window is long enough (e.g., such that the average rate $\overline \beta$ of incoming tweets should yield approximately similar values when time-averaged over disjoint time windows of length $W$). Similarly, the expected number of retweets by the user during the time window can be written as in Eq.~(\ref{Nretweets}), but with the Poisson rate $\beta_{j k}$ replaced by its time-averaged value. The calculations of Eq.~(\ref{xijkB}) then proceed as before, leading to the conclusion that the branching number limits to the critical value of one as $\mu \to 0$, which implies that non-Poisson user activity rates (or burstiness) will not affect the criticality of the model, which is a long-time (i.e., ages of memes limit to infinity) characteristic. Of course, the short-term behaviour of the model  (such as the small-$a$ behaviour in panels B--E of Fig.~\ref{fig4}) would be affected by introducing burstiness; incorporating such realistic features into the model is left as a challenge for further work.
As a final comment on this topic, we note that the agreement (in Sec.~\ref{resultsdata}) of our theoretical results with real data of a spreading process
for which users' activity rates are not constant also provides indirect evidence that the phenomenology
discussed is robust to the details of user activity burstiness.

The heuristic calculation  of the branching number considered at the end of Sec.~\ref{criticality} also
 offers a clue as to how the model can be extended to the spreading of information on undirected
  social networks (as opposed to the directed networks that we focus on in this paper). Of particular interest is the spreading of app-adoption on Facebook, for which data was analyzed in Ref.~\cite{Onnela10} and a computational model was introduced in Ref.~\cite{GleesonPNAS14}. If the Facebook update messages that inform all friends of  user $A$ that she has installed a particular app are considered to be the memes in a version of our model, then the arguments of Sec.~\ref{criticality} need only slight modifications. The total number of update messages received in the stream of a user with $k$ Facebook friends is linear in $k$ (i.e., the $j$ in the denominator of Eq.~(\ref{Pchosen}) is replaced by $k$), while the expected number of friends who would be interested in  user $A$'s adoption of the app is $\lambda (k-1)$ (since one friend out of $k$ must have adopted before
   $A$ in order to have spread the message to her). Following very similar steps to calculate the expected number $\xi$ of children of a meme---see the calculations leading to Eq.~(\ref{xijkB})---we find that
\begin{align}
\xi_\text{undirected} &\to \sum_k\frac{k}{z}p_k \frac{\beta_k \lambda (k-1)}{k \overline{\beta}\lambda}  \quad\text{ as }\mu \to 0 \nonumber \\
& = 1-\frac{1}{z}.
\end{align}
Although this branching number is less than one, the mean number $z$ of friends on Facebook is large (e.g., Ref.~\cite{Ugander11} calculated $z\approx 190$) so that $\xi_\text{undirected}$ is in fact very close to unity, implying that the information spread process is close to criticality. Such a near-critical branching process was hypothesized in Ref.~\cite{GleesonPNAS14} to explain to observed fat-tailed
distributions of app popularity in Facebook data and the temporal characteristics of the adoption behaviour.
 The cascade sizes for other forms of ``meme'' spreading on Facebook have also been observed to have fat-tailed distributions \cite{Adamic14}. Other undirected networks to which the model should be applicable include YouTube \cite{Szabo10} and Digg \cite{Wu07,Lerman12}.

Finally, our focus here has been on the statistical physics of the model, but for completeness we should note the difficulties inherent in applying the model to data sets where memes may not be as simple to recognize and track as hashtags are. In \cite{Kuhn14} for example, the process of extracting memes (representing popular scientific terms) from data (citation archives of scientific publications) is explained in detail, and considerable such effort will generally be required to identify and track the memes to which this null model might be applied.

A related question is whether the popularity of online memes has any implications in terms of mass social movements in the offline world. This is a complex question that lies beyond the scope of this paper, but we note that Fig.~3 of \cite{gonzalez13} shows that the usage of hashtags related to the 15M Spanish protest movement was
found to be closely correlated with the number of protest-related headlines in newspapers, at least during the main activity of the protests. This indicates that online social spreading phenomena can, at least in some cases, give useful information about real-world social movements and activism.
}

\section{Discussion} \label{Discussion}
The extremely wide range of popularities achieved by items on social media poses many challenges for complex systems researchers. These include the identification of  the causes \cite{Coscia14} and structural features \cite{Goel15} of ``viral'' propagation, and the prediction of future spreading based on the content or the early-time growth of memes \cite{Cheng14,Weng13,Miotto14,Thij14}, each of which are important in the design of more efficient systems to spread information (e.g. in case of emergency). We argue that null models are fundamentally important in this quest---and complement more data-driven approaches---as they demonstrate, for example, that extreme popularity can arise purely because of random fluctuations in the competition between memes for user attention. While the content of a meme may well be an important factor in its popularity (or predictability \cite{Miotto14}), definitive statements about the significance of such factors should be referenced to an appropriate null model.

In this paper we  have introduced and analyzed  a null model of meme spreading that is analytically tractable, yet realistic enough to reproduce several characteristic features of empirical data.
The model is sufficiently general to incorporate heterogeneous user activity rates and a joint distribution $p_{jk}$ of the number of users followed $j$ and the number of followers $k$, as well as a memory-time distribution $\Phi$ that gives non-Markovian dynamics. The competition-induced criticality phenomenon identified in a zero-memory model in Ref.~\cite{GleesonPRL14} is found to be robust to the generalizations, giving power-law popularity distributions with characteristic time-dependence similar to data from social spreading phenomena
{ (and see Sec.~\ref{limitations} for a discussion of further possible extensions of the model).}

The analytical tractability enables fast fitting of the model to data, as demonstrated in Sec.~\ref{resultsdata} with hashtag data from Twitter. We find that a simplified version of the model where users all have the same activity rate can be fitted to some, but not all, aspects of the data (see Fig.~\ref{fig6}). The aim of a null model is not to perfectly reproduce every aspect of a dataset, but rather to help identify which features of the data can be reproduced using relatively simple models, and so to highlight aspects where more detailed modelling (or, perhaps, factors entirely outside the model) are required to match to data. In this respect, the null model highlights the fact that heterogeneity in activity rates is vital to accurately capturing the $q_1(a)$ curve (compare Figs.~\ref{fig6}D and \ref{figS2}D), even though the time-dependence of the bulk of the popularity distribution may be described reasonably well  by a model with homogeneous activity rates (Fig.~\ref{fig6}A-C).

{ As noted in the Introduction, and expanded upon in Sec.~\ref{limitations}, our definition of ``memes'' is sufficiently general to enable the model to be applied (with minor changes) not just to the spreading of hashtags or URLs on Twitter, but also to the adoption of apps on Facebook, the popularity of videos on YouTube, and to the broad range of imitation-driven spreading dynamics.}
We anticipate that the analytical results and potential for fast fitting to data will make this null model a useful tool for further work, and we hope it will contribute to the ongoing investigation  of the entangled effects of memory, network structure, and competition on social spreading phenomena.

\section*{Acknowledgements}
The authors acknowledge helpful feedback from Davide Cellai, Timothy Duff, Rick Durrett, Freja Elbro, Ali Faqeeh, Peter Fennell, Kristina Lerman, David O'Sullivan, Mason Porter, and  Jonathan Ward. This work was supported by Science Foundation Ireland (grant numbers 11/PI/1026, 12/IA/1683, and 09/SRC/E1780, J.P.G. and K.P.O'S.) and by the European Commission through FET-Proactive projects PLEXMATH (FP7-ICT-2011-8; grant number 317614, J.P.G and Y.M.) and MULTIPLEX (FP7-ICT-2011-8; grant 317532, Y.M.). R.A.B. and Y.M. also acknowledge support from MINECO (Grant FIS2011-25167) and Comunidad de Arag{\'o}n (Spain; grant  FENOL). We acknowledge the SFI/HEA Irish Centre for High-End Computing (ICHEC) for the provision of computational facilities. The data used in this paper is available for download from www.ul.ie/gleesonj/twitter15M.

\appendix
\numberwithin{equation}{section}
\section{Calculation of the large-$a$, large-$n$ asymptotics of popularity distribution}\label{AppA}

In this Appendix we consider how the popularity distribution $q_n(a)$ behaves for large, but finite, ages. To highlight the effect of the out-degree distribution $p_k$ upon the results we here restrict
our analysis to the case $\beta_{jk}\equiv 1$, $p_{jk}=\delta_{j,z}p_k$. Taking the $\mu\to 0$ limit, Eq.~(\ref{e19}) becomes
\begin{align}
G(a;x)&=\sum_{ k}p_{k} \int_0^\infty d\ell\, \lambda z e^{-\lambda z \ell} \times\nonumber\\
 &\times\exp\left[-\int_{0}^{\min(\ell,a)} d\tilde{r}\, \int_{0}^{a-\tilde{r}} d\tilde{\t}\,\Phi(a-\tilde{r}-\tilde{\t}) (1- x\left[1-\lambda+\lambda G(\tilde{\t};x)\right]^k)\right].\label{e19b}
\end{align}
Writing $x=1-w$ and $G(a;x)=1-\phi(a;w)$, we observe that the argument of the exponential function vanishes when $w=0$ and $\phi=0$, and so we consider the small-$w$, small-$\phi$ asymptotic behavior by expanding the exponential term to first order in its argument:
\begin{equation}
\phi(a;w)\approx \int_0^\infty d\ell\, \lambda z e^{-\lambda z \ell} \int_{0}^{\min(\ell,a)} d\tilde{r}\, \int_{0}^{a-\tilde{r}} d\tilde{\t}\,\Phi(a-\tilde{r}-\tilde{\t}) (1- (1-w)\sum_k p_k \left[1-\lambda\phi(\tilde{\t};w)\right]^k).\label{onetermexpn}
\end{equation}
We note that retaining only the first-order term in the expansion of the exponential is an approximation. We will estimate the accuracy of this ``one-term expansion'' by comparing the infinite-age limit determined under the approximation with the corresponding exact values as given in Sec.~\ref{sec:S1.6}.

\begin{figure}
\centering
\epsfig{figure=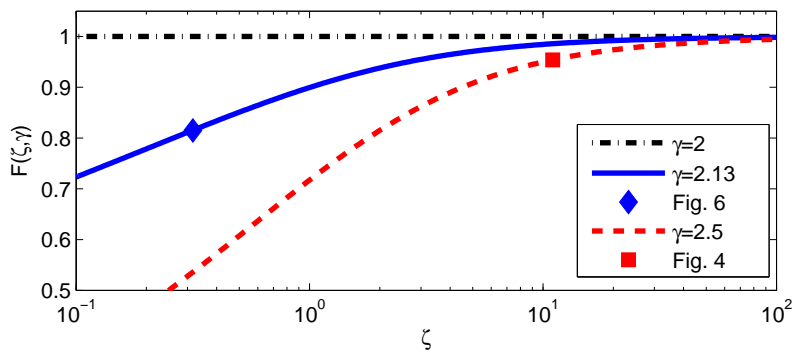,width=12cm}
\caption{The function $F(\zeta,\gamma)$, as defined in Eq.~(\ref{Fdefn}), for values of $\gamma$ close to 2. The highlighted points are the parameter values that are relevant to Fig.~\ref{fig4} ($\lambda z = 11$, $\gamma=2.5$) and to Fig.~\ref{fig6} ($\lambda z = 0.32$, $\gamma=2.13$). In all cases of interest the values of $F$ are close to 1, and so we conclude that the one-term expansion used in Eq.~(\ref{onetermexpn}) gives a good estimate of the exact steady-state solution.}
\label{figFplot}
\end{figure}

For the case of a scale-free out-degree distribution with $p_k\sim D \,k^{-\gamma}$ as $k\to \infty$, and $\gamma$ in the range $2<\gamma<3$,  the asymptotic
form of the summation term  in Eq.~(\ref{onetermexpn}) is given by \cite{GleesonPRL14}
\begin{equation}
1-(1-w)\sum_{k}p_k\left[1-\lambda\phi\right]^k \sim \lambda z \phi - C \phi^{\gamma-1}+w +o(w,\phi) \quad\text{ as }w\to0,\phi\to0,
\end{equation}
with the constant $C$ given by $C=\lambda^{\gamma-1}D \Gamma(1-\gamma)$. Applying the integral-swapping trick of Eq.~(\ref{swap}) allows the right hand side of Eq.~(\ref{onetermexpn}) to be expressed as a double convolution integral. Laplace transforming then yields
\begin{equation}
\hat \phi(s;w) = \frac{1}{\lambda z +s}\hat\Phi(s) \mathcal{L}\left[ \lambda z \phi - C \phi^{\gamma-1}+w\right], \label{phia1}
\end{equation}
where $\mathcal{L}$ denotes the Laplace transform operation applied to the term in square brackets.
In the $a\to \infty$ limit, this equation is satisfied by the steady-state solution
\begin{equation}
\phi(\infty;w) = C^{-\frac{1}{\gamma-1}} w^{\frac{1}{\gamma-1}},\label{S51}
\end{equation}
as can be verified using the final value theorem for Laplace transforms.
We note that the corresponding expression for $\phi(\infty;w)$ as calculated from the steady state Eq.~(\ref{e24}) has a additional multiplicative factor of $F(\lambda z,\gamma)$  that is absent in Eq.~(\ref{S51}), where the function $F(\zeta,\gamma)$ is defined by
\begin{equation}
F(\zeta,\gamma) = \left[\zeta^2\sum_{n=1}^\infty \frac{n^{\gamma-1}}{(\zeta+1)^{n+1}}\right]^{-\frac{1}{\gamma-1}}, \label{Fdefn}
\end{equation}
see Fig.~\ref{figFplot}.
If $\lambda z \gg 1$, then $F(\lambda z,\gamma)\approx 1$ and the one-term expansion gives results that are very close to the exact values (at least in the infinite-age limit $a\to\infty$).
Moreover, even if $\lambda z$ is not large (e.g., $\lambda z=0.32$ for the model fit to Twitter hashtags data in Sec.~\ref{resultsdata}), the values of $F(\lambda z,\gamma)$ can still be close to unity if $\gamma$ is sufficiently close to 2.

To consider small deviations from the steady state, we define $g(a;w)$ 
 by
\begin{equation}
\phi(a;w)=\phi(\infty;w)\left( 1-g(a;w)\right) \label{phia2}
\end{equation}
with $g(a;w)\to0$ as $a\to\infty$. Assuming that $g$ is sufficiently small to allow the use of the linearizing  approximation
\begin{equation}
(1-g)^{\gamma-1} \approx 1-(\gamma-1)g,
\end{equation}
 Eq.~(\ref{phia1}) can be solved for the Laplace transform of $g$:
\begin{equation}
\hat g (s;w) = \frac{1}{s}\frac{s+\lambda z - \lambda z \hat\Phi(s)}{s+\lambda z - \lambda z \hat\Phi(s) + (\gamma-1)C^\frac{1}{\gamma-1}w^{\frac{\gamma-2}{\gamma-1}}\hat\Phi(s)}. \label{phia3}
\end{equation}
The Laplace transform of $\phi$ then follows from Eq.~(\ref{phia2}) and a similar asymptotic analysis of Eq.~(\ref{e21}) yields
\begin{equation}
\hat H(s;1-w) = \frac{1}{s}-\frac{\lambda z(s+\lambda z +\hat \Phi(s))}{s+\lambda z}\hat\phi(s;w)\label{eS56}
\end{equation}
Substituting from Eqs.~(\ref{phia2}) and (\ref{phia3}) results in
\begin{equation}
\hat H(s;x) = \frac{1}{s}\left[ 1- \frac{\lambda z\left(s+\lambda z+\hat \Phi(s)\right)(\gamma-1)(1-x)\hat \Phi(s)}{(s+\lambda z)\left(s+\lambda z - \lambda z \hat\Phi(s)+(\gamma-1)\lambda D^\frac{1}{\gamma-1}\left[\Gamma(1-\gamma)\right]^\frac{1}{\gamma-1}(1-x)^\frac{\gamma-2}{\gamma-1}
\hat\Phi(s)\right)}\right].\label{H1}
\end{equation}

A similar analysis can be performed in the case where the out-degree distribution $p_k$ has finite second moment. We again utilize a one-term expansion similar to Eq.~(\ref{onetermexpn}), but we can also retain a non-vanishing innovation probability $\mu$ in this case. The one-term expansion can be shown to be accurate when $\lambda z \gg 1$; this condition is obeyed in all relevant cases we examine. The resulting large-$a$ asymptotics for the generating function $H(a;x)$ are found by inverting the following Laplace transform:
\begin{align}
&\hat H(s;1-w) = \frac{1}{s}-\phi(\infty;w)\times&\nonumber\\
&\hspace{0cm}\times\frac{(1-\mu)\lambda z(s+\lambda z+\mu +\hat \Phi(s))}{s(s+\lambda z+\mu)}\left[ \frac{2(1-\mu)\frac{w}{\phi(\infty;w)} \hat \Phi(s) - \mu(\lambda z+1)\hat \Phi(s)}{s+\lambda z +\mu - \left(\lambda z(1+\mu)+2\mu\right)\hat \Phi(s) + 2(1-\mu)\frac{w}{\phi(\infty;w)}\hat \Phi(s)}\right], \label{Hasym2}
\end{align}
with $\phi(\infty;w)$ given by
\begin{equation}
\phi(\infty;w) = \frac{-\mu(\lambda z +1)+\sqrt{\mu^2(\lambda z+1)^2+2\lambda^2(1-\mu)^2\left(\left<k^2\right>-z\right)w}}{\lambda^2(1-\mu)\left(\left<k^2\right>-z\right)}.
\end{equation}

 \section{Numerical inversion of Laplace transforms and PGFs}\label{sec:inversion}
 Many of our  results for the popularity distribution $q_n(a)$ are expressed in terms of the corresponding PGF $H(a;x)$. As in \cite{GleesonPRL14}, we use the Fast Fourier Transform method of \cite{Cavers78,Newman01,Marder07,Abate92} to numerically invert the PGF at a fixed age $a$ to produce, for example, the model distributions in Figs.~\ref{fig6} and \ref{figS2}; see Sec.~S2 of \cite{GleesonPRL14} for further details and links to Octave/Matlab code for implementing the PGF inversion.

 The results of the model for the age-dependence of several quantities are expressed in terms of Laplace transforms. To numerically invert the Laplace transforms we use the efficient Talbot algorithm \cite{Talbot79}, in its simplified version described in Sec.~6 of \cite{Abate06}. The Talbot algorithm is based on a numerical evaluation of the Bromwich (Laplace inversion) integral, using a cleverly-chosen deformation of the contour in the complex-$s$ plane. The Laplace inversion of $\hat H(s;x)$ to obtain $H(a;x)$ at a desired age $a$, for example, can be quickly computed using the $2 M_L-1$ weights $\gamma_k$ and nodes $\delta_k$ defined by \cite{Abate04}
 \begin{align}
 \delta_0&= \frac{2M_L}{5}, \quad \delta_k=\frac{2k\pi}{5}(\cot(k\pi/M_L)+i)\quad\text{ for }-M_L+1\le k\le M_L-1,\nonumber\\
 \gamma_0&=\frac{1}{2}e^{\delta_0},\nonumber\\
 \gamma_k&=[1+i(k\pi/M_L)(1+[\cot(k\pi/M_L)^2])-i\cot(k\pi/M_L)]e^{\delta_k}\quad\text{ for }-M_L+1\le k\le M_L-1,
 \end{align}
 (where $i=\sqrt{-1}$) by calculating
 the sum
\begin{equation}
H(a;x) = \frac{1}{5a}\left[ \gamma_0 \hat H\left(\frac{\delta_0}{a};x\right)+\sum_{k=-M_L+1}^{M_L-1} \gamma_k \hat H\left(\frac{\delta_k}{a};x\right)\right].
 \end{equation}
 In practice, the precision of the Talbot algorithm is very high, and only relatively small values of $M_L$ are required to obtain accurate results; we used $M_L=25$ in the examples shown.

\section{Model of heterogeneous activity rates}\label{AppC}
 In the data analysis of Fig.~6(a) of \cite{Hodas13}, the average activity rate (as measured by the number of tweets by a user in a fixed time period) is found to grow approximately linearly with the number of followers $k$ of that user, for $k$ from 0 to about 100. Then, for $k$ values from about 100 up to the maximum shown in the plot ($k=10^3$), the activity rate grows as a more slowly increasing linear function of $k$. We model these characteristics (which are also seen in other studies, e.g., \cite{Myers14}), using a piecewise-linear and continuous  function of $k$, assuming a saturation of activity at very high $k$, as follows:
\begin{equation}
\beta_k \propto \left\{ \begin{array}{cc}
0.35 k & \text{ if }k<100,\\
35+0.044(k-100) & \text{ if }100\le k<10^4.\\
470.6 & \text{ if }k\ge10^4,
\end{array}
\right. \label{betak}
\end{equation}
where the values are chosen to closely match the linear growth rates in Fig.~6(a) of \cite{Hodas13}, snd the constant of proportionality being set by the condition $\sum_k \beta_k p_k = 1$.

\bibliography{CIC2,compete_bib}

\begin{thebibliography}{10}

\bibitem{Bakshy11}
Eytan Bakshy, Jake~M Hofman, Winter~A Mason, and Duncan~J Watts.
\newblock Everyone's an influencer: quantifying influence on {T}witter.
\newblock In {\em Proceedings of the Fourth ACM International Conference on Web
  Search and Data Mining}, pages 65--74. ACM, 2011.

\bibitem{Lerman12}
Kristina Lerman, Rumi Ghosh, and Tawan Surachawala.
\newblock Social contagion: An empirical study of information spread on {D}igg
  and {T}witter follower graphs.
\newblock {\em arXiv preprint arXiv:1202.3162}, 2012.

\bibitem{Weng12}
Lillian Weng, Alessandro Flammini, Alessandro Vespignani, and Fillipo Menczer.
\newblock Competition among memes in a world with limited attention.
\newblock {\em Scientific Reports}, 2, 2012.

\bibitem{Cheng14}
Justin Cheng, Lada Adamic, P~Alex Dow, Jon~Michael Kleinberg, and Jure
  Leskovec.
\newblock Can cascades be predicted?
\newblock In {\em Proceedings of the 23rd International Conference on World
  Wide Web}, pages 925--936. ACM, 2014.

\bibitem{GleesonPNAS14}
James~P Gleeson, Davide Cellai, Jukka-Pekka Onnela, Mason~A Porter, and Felix
  Reed-Tsochas.
\newblock A simple generative model of collective online behavior.
\newblock {\em Proceedings of the National Academy of Sciences},
  111(29):10411--10415, 2014.

\bibitem{Castellano09}
Claudio Castellano, Santo Fortunato, and Vittorio Loreto.
\newblock Statistical physics of social dynamics.
\newblock {\em Reviews of Modern Physics}, 81(2):591, 2009.

\bibitem{Leskovec07}
Jure Leskovec, Lada~A Adamic, and Bernardo~A Huberman.
\newblock The dynamics of viral marketing.
\newblock {\em ACM Transactions on the Web (TWEB)}, 1(1):5, 2007.

\bibitem{PastorSatorras01}
Romualdo Pastor-Satorras and Alessandro Vespignani.
\newblock Epidemic spreading in scale-free networks.
\newblock {\em Physical Review Letters}, 86(14):3200, 2001.

\bibitem{diseasereview}
Romualdo Pastor-Satorras, Claudio Castellano, Piet Van~Mieghem, and Alessandro
  Vespignani.
\newblock Epidemic processes in complex networks.
\newblock {\em Reviews of Modern Physics}, 87(3):925, 2015.

\bibitem{Onnela10}
Jukka-Pekka Onnela and Felix Reed-Tsochas.
\newblock Spontaneous emergence of social influence in online systems.
\newblock {\em Proceedings of the National Academy of Sciences},
  107(43):18375--18380, 2010.

\bibitem{Miotto}
Jos{\'e}~M Miotto and Eduardo~G Altmann.
\newblock Predictability of extreme events in social media.
\newblock {\em PLOS ONE}, 9(11):e111506, 2014.

\bibitem{Dawkins}
Richard Dawkins.
\newblock {\em The selfish gene}.
\newblock Oxford University Press, 2006.

\bibitem{memedefinition}
\url{http://oxforddictionaries.com/definition/english/meme}.

\bibitem{Adamic14}
Lada~A Adamic, Thomas~M Lento, Eytan Adar, and Pauline~C Ng.
\newblock Information evolution in social networks.
\newblock {\em arXiv preprint arXiv:1402.6792}, 2014.

\bibitem{Leskovec09}
Jure Leskovec, Lars Backstrom, and Jon Kleinberg.
\newblock Meme-tracking and the dynamics of the news cycle.
\newblock In {\em Proceedings of the 15th ACM SIGKDD international conference
  on Knowledge discovery and data mining}, pages 497--506. ACM, 2009.

\bibitem{Weng13}
Lilian Weng, Filippo Menczer, and Yong-Yeol Ahn.
\newblock Virality prediction and community structure in social networks.
\newblock {\em Scientific Reports}, 3, 2013.

\bibitem{Bentley04}
R~Alexander Bentley, Matthew~W Hahn, and Stephen~J Shennan.
\newblock Random drift and culture change.
\newblock {\em Proceedings of the Royal Society of London B: Biological
  Sciences}, 271(1547):1443--1450, 2004.

\bibitem{Herzog04}
Harold~A Herzog, R~Alexander Bentley, and Matthew~W Hahn.
\newblock Random drift and large shifts in popularity of dog breeds.
\newblock {\em Proceedings of the Royal Society of London B: Biological
  Sciences}, 271(Suppl 5):S353--S356, 2004.

\bibitem{Simkin07}
Mikhail~V Simkin and Vwani~P Roychowdhury.
\newblock A mathematical theory of citing.
\newblock {\em Journal of the American Society for Information Science and
  Technology}, 58(11):1661--1673, 2007.

\bibitem{Redner98}
Sidney Redner.
\newblock How popular is your paper? an empirical study of the citation
  distribution.
\newblock {\em The European Physical Journal B-Condensed Matter and Complex
  Systems}, 4(2):131--134, 1998.

\bibitem{Kuhn14}
Tobias Kuhn, Matja{\v{z}} Perc, and Dirk Helbing.
\newblock Inheritance patterns in citation networks reveal scientific memes.
\newblock {\em Physical Review X}, 4(4):041036, 2014.

\bibitem{Price76}
Derek de~Solla Price.
\newblock A general theory of bibliometric and other cumulative advantage
  processes.
\newblock {\em Journal of the American society for Information Science},
  27(5):292--306, 1976.

\bibitem{Newman05}
M.~E.~J. Newman.
\newblock Power laws, {P}areto distributions and {Z}ipf's law.
\newblock {\em Contemporary Physics}, 46(5):323--351, 2005.

\bibitem{Perc14}
Matja{\v{z}} Perc.
\newblock The {M}atthew effect in empirical data.
\newblock {\em Journal of The Royal Society Interface}, 11(98):20140378, 2014.

\bibitem{Barabasi99}
Albert-L{\'a}szl{\'o} Barab{\'a}si and R{\'e}ka Albert.
\newblock Emergence of scaling in random networks.
\newblock {\em Science}, 286(5439):509--512, 1999.

\bibitem{Sethna01}
James~P Sethna, Karin~A Dahmen, and Christopher~R Myers.
\newblock Crackling noise.
\newblock {\em Nature}, 410(6825):242--250, 2001.

\bibitem{Stanleybook}
H~Eugene Stanley.
\newblock {\em Introduction to phase transitions and critical phenomena}.
\newblock Oxford University Press, 1987.

\bibitem{Bakbook}
P.~Bak.
\newblock {\em How nature works: the science of self-organized criticality}.
\newblock Springer, 1999.

\bibitem{Barabasi05}
Albert-Laszlo Barabasi.
\newblock The origin of bursts and heavy tails in human dynamics.
\newblock {\em Nature}, 435(7039):207--211, 2005.

\bibitem{Malmgren08}
R~Dean Malmgren, Daniel~B Stouffer, Adilson~E Motter, and Lu{\'\i}s~AN Amaral.
\newblock A {P}oissonian explanation for heavy tails in e-mail communication.
\newblock {\em Proceedings of the National Academy of Sciences},
  105(47):18153--18158, 2008.

\bibitem{Karsai11}
M{\'a}rton Karsai, Mikko Kivel{\"a}, Raj~Kumar Pan, Kimmo Kaski, J{\'a}nos
  Kert{\'e}sz, A-L Barab{\'a}si, and Jari Saram{\"a}ki.
\newblock Small but slow world: How network topology and burstiness slow down
  spreading.
\newblock {\em Physical Review E}, 83(2):025102, 2011.

\bibitem{Delvenne15}
Jean-Charles Delvenne, Renaud Lambiotte, and Luis~EC Rocha.
\newblock Diffusion on networked systems is a question of time or structure.
\newblock {\em Nature Communications}, 6, 2015.

\bibitem{JoPRX14}
Hang-Hyun Jo, Juan~I Perotti, Kimmo Kaski, and J{\'a}nos Kert{\'e}sz.
\newblock Analytically solvable model of spreading dynamics with
  non-{P}oissonian processes.
\newblock {\em Physical Review X}, 4(1):011041, 2014.

\bibitem{Iribarren09}
Jos{\'e}~Luis Iribarren and Esteban Moro.
\newblock Impact of human activity patterns on the dynamics of information
  diffusion.
\newblock {\em Physical Review Letters}, 103(3):038702, 2009.

\bibitem{Iribarren11}
Jos{\'e}~Luis Iribarren and Esteban Moro.
\newblock Branching dynamics of viral information spreading.
\newblock {\em Physical Review E}, 84(4):046116, 2011.

\bibitem{Bentley11}
R~Alexander Bentley, Paul Ormerod, and Michael Batty.
\newblock Evolving social influence in large populations.
\newblock {\em Behavioral Ecology and Sociobiology}, 65(3):537--546, 2011.

\bibitem{GleesonPRL14}
James~P Gleeson, Jonathan~A Ward, Kevin~P O'Sullivan, and William~T Lee.
\newblock Competition-induced criticality in a model of meme popularity.
\newblock {\em Physical Review Letters}, 112(4):048701, 2014.

\bibitem{Hodas13}
Nathan~O Hodas, Farshad Kooti, and Kristina Lerman.
\newblock Friendship paradox redux: Your friends are more interesting than you.
\newblock {\em arXiv preprint arXiv:1304.3480}, 2013.

\bibitem{Pinto11}
Oscar~A Pinto and Miguel~A Mu{\~n}oz.
\newblock Quasi-neutral theory of epidemic outbreaks.
\newblock {\em PLOS ONE}, 6(7):e21946, 2011.

\bibitem{Kimurabook}
Motoo Kimura.
\newblock {\em The neutral theory of molecular evolution}.
\newblock Cambridge University Press, 1984.

\bibitem{Bianconi01}
Ginestra Bianconi and A-L Barab{\'a}si.
\newblock Competition and multiscaling in evolving networks.
\newblock {\em EPL (Europhysics Letters)}, 54(4):436, 2001.

\bibitem{Simkin11}
Mikhail~V Simkin and Vwani~P Roychowdhury.
\newblock Re-inventing {W}illis.
\newblock {\em Physics Reports}, 502(1):1--35, 2011.

\bibitem{Simon55}
Herbert~A Simon.
\newblock On a class of skew distribution functions.
\newblock {\em Biometrika}, 42(3/4):425--440, 1955.

\bibitem{Harrisbook}
Theodore~E Harris.
\newblock {\em The theory of branching processes}.
\newblock Courier Corporation, 2002.

\bibitem{Wilfbook}
Herbert~S Wilf.
\newblock {\em generatingfunctionology}.
\newblock Elsevier, 2013.

\bibitem{Newman01}
Mark~EJ Newman, Steven~H Strogatz, and Duncan~J Watts.
\newblock Random graphs with arbitrary degree distributions and their
  applications.
\newblock {\em Physical Review E}, 64(2):026118, 2001.

\bibitem{Goh03}
K-I Goh, D-S Lee, B~Kahng, and D~Kim.
\newblock Sandpile on scale-free networks.
\newblock {\em Physical Review Letters}, 91(14):148701, 2003.

\bibitem{SchwartzCohen}
N~Schwartz, R~Cohen, D~Ben-Avraham, A-L Barab{\'a}si, and S~Havlin.
\newblock Percolation in directed scale-free networks.
\newblock {\em Physical Review E}, 66(1):015104, 2002.

\bibitem{Adami02}
Christoph Adami and Johan Chu.
\newblock Critical and near-critical branching processes.
\newblock {\em Physical Review E}, 66(1):011907, 2002.

\bibitem{Zapperi95}
Stefano Zapperi, Kent~B{\ae}kgaard Lauritsen, and H~Eugene Stanley.
\newblock Self-organized branching processes: mean-field theory for avalanches.
\newblock {\em Physical Review Letters}, 75(22):4071, 1995.

\bibitem{Hodas12}
Nathan~O Hodas and Kristina Lerman.
\newblock How visibility and divided attention constrain social contagion.
\newblock In {\em Privacy, Security, Risk and Trust (PASSAT), 2012
  International Conference on and 2012 International Confernece on Social
  Computing (SocialCom)}, pages 249--257. IEEE, 2012.

\bibitem{Newman09}
M.~E.~J. Newman.
\newblock The first-mover advantage in scientific publication.
\newblock {\em EPL (Europhysics Letters)}, 86(6):68001, 2009.

\bibitem{Vazquez07}
Alexei Vazquez, Balazs Racz, Andras Lukacs, and Albert-Laszlo Barabasi.
\newblock Impact of non-{P}oissonian activity patterns on spreading processes.
\newblock {\em Physical Review Letters}, 98(15):158702, 2007.

\bibitem{Min11}
Byungjoon Min, K-I Goh, and Alexei Vazquez.
\newblock Spreading dynamics following bursty human activity patterns.
\newblock {\em Physical Review E}, 83(3):036102, 2011.

\bibitem{TemporalNetworks}
Petter Holme and Jari Saram{\"a}ki.
\newblock Temporal networks.
\newblock {\em Physics Reports}, 519(3):97--125, 2012.

\bibitem{Hoffmann12}
Till Hoffmann, Mason~A Porter, and Renaud Lambiotte.
\newblock Generalized master equations for non-{P}oisson dynamics on networks.
\newblock {\em Physical Review E}, 86(4):046102, 2012.

\bibitem{Boguna14}
Marian Bogun{\'a}, Luis~F Lafuerza, Ra{\'u}l Toral, and M~{\'A}ngeles Serrano.
\newblock Simulating non-{M}arkovian stochastic processes.
\newblock {\em Physical Review E}, 90(4):042108, 2014.

\bibitem{VanMieghem13}
P~Van~Mieghem and R~Van~de Bovenkamp.
\newblock Non-{M}arkovian infection spread dramatically alters the
  susceptible-infected-susceptible epidemic threshold in networks.
\newblock {\em Physical Review Letters}, 110(10):108701, 2013.

\bibitem{Kivela14}
Mikko Kivel{\"a} and Mason~A Porter.
\newblock Estimating interevent time distributions from finite observation
  periods in communication networks.
\newblock {\em Physical Review E}, 92(5):052813, 2015.

\bibitem{AthreyaNeybook}
Krishna~B Athreya and Peter~E Ney.
\newblock {\em Branching processes}, volume 196.
\newblock Springer Science \& Business Media, 2012.

\bibitem{Doerr13}
Christian Doerr, Norbert Blenn, and Piet Van~Mieghem.
\newblock Lognormal infection times of online information spread.
\newblock {\em PLOS ONE}, 8(5):e64349, 2013.

\bibitem{Banos13}
Raquel~A Ba{\~n}os, Javier Borge-Holthoefer, and Yamir Moreno.
\newblock The role of hidden influentials in the diffusion of online
  information cascades.
\newblock {\em EPJ Data Science}, 2(1):1--16, 2013.

\bibitem{Szabo10}
Gabor Szabo and Bernardo~A Huberman.
\newblock Predicting the popularity of online content.
\newblock {\em Communications of the ACM}, 53(8):80--88, 2010.

\bibitem{Borge-Holthoefer11}
Javier Borge-Holthoefer, Alejandro Rivero, I{\~n}igo Garc{\'\i}a, Elisa
  Cauh{\'e}, Alfredo Ferrer, Dar{\'\i}o Ferrer, David Francos, David
  I{\~n}iguez, Mar{\'\i}a~Pilar P{\'e}rez, Gonzalo Ruiz, et~al.
\newblock Structural and dynamical patterns on online social networks: the
  {S}panish {M}ay 15th movement as a case study.
\newblock {\em PLOS ONE}, 6(8):e23883, 2011.

\bibitem{Gonzalez-Bailon11}
Sandra Gonz{\'a}lez-Bail{\'o}n, Javier Borge-Holthoefer, Alejandro Rivero, and
  Yamir Moreno.
\newblock The dynamics of protest recruitment through an online network.
\newblock {\em Scientific Reports}, 1, 2011.

\bibitem{Clauset09}
Aaron Clauset, Cosma~Rohilla Shalizi, and Mark~EJ Newman.
\newblock Power-law distributions in empirical data.
\newblock {\em SIAM Review}, 51(4):661--703, 2009.

\bibitem{Boguna05}
Mari{\'a}n Bogu{\~n}{\'a} and M~{\'A}ngeles Serrano.
\newblock Generalized percolation in random directed networks.
\newblock {\em Physical Review E}, 72(1):016106, 2005.

\bibitem{Hurd16}
TR~Hurd.
\newblock The construction and properties of assortative configuration graphs.
\newblock {\em arXiv preprint arXiv:1512.03084}, 2015.

\bibitem{Kwak10}
Haewoon Kwak, Changhyun Lee, Hosung Park, and Sue Moon.
\newblock What is {T}witter, a social network or a news media?
\newblock In {\em Proceedings of the 19th International Conference on World
  Wide Web}, pages 591--600. ACM, 2010.

\bibitem{Melnik11}
Sergey Melnik, Adam Hackett, Mason~A Porter, Peter~J Mucha, and James~P.
  Gleeson.
\newblock The unreasonable effectiveness of tree-based theory for networks with
  clustering.
\newblock {\em Physical Review E}, 83(3):036112, 2011.

\bibitem{Bentleybook}
Alex Bentley, Mark Earls, Michael~J O'Brien, and John Maeda.
\newblock {\em I'll have what she's having: Mapping social behavior}.
\newblock MIT Press, 2011.

\bibitem{Salganik06}
Matthew~J Salganik, Peter~Sheridan Dodds, and Duncan~J Watts.
\newblock Experimental study of inequality and unpredictability in an
  artificial cultural market.
\newblock {\em Science}, 311(5762):854--856, 2006.

\bibitem{Perra12}
Nicola Perra, Bruno Gon{\c{c}}alves, Romualdo Pastor-Satorras, and Alessandro
  Vespignani.
\newblock Activity driven modeling of time varying networks.
\newblock {\em Scientific Reports}, 2, 2012.

\bibitem{Ugander11}
Johan Ugander, Brian Karrer, Lars Backstrom, and Cameron Marlow.
\newblock The anatomy of the {F}acebook social graph.
\newblock {\em arXiv preprint arXiv:1111.4503}, 2011.

\bibitem{Wu07}
Fang Wu and Bernardo~A Huberman.
\newblock Novelty and collective attention.
\newblock {\em Proceedings of the National Academy of Sciences},
  104(45):17599--17601, 2007.

\bibitem{gonzalez13}
Sandra Gonz{\'a}lez-Bail{\'o}n, Javier Borge-Holthoefer, and Yamir Moreno.
\newblock Broadcasters and hidden influentials in online protest diffusion.
\newblock {\em American Behavioral Scientist}, 57(7):943--965, 2013.

\bibitem{Coscia14}
Michele Coscia.
\newblock Average is boring: How similarity kills a meme's success.
\newblock {\em Scientific Reports}, 4, 2014.

\bibitem{Goel15}
Sharad Goel, Ashton Anderson, Jake Hofman, and Duncan~J Watts.
\newblock The structural virality of online diffusion.
\newblock {\em Management Science}, 62(1):180--196, 2015.

\bibitem{Miotto14}
Jos{\'e}~M Miotto and Eduardo~G Altmann.
\newblock Predictability of extreme events in social media.
\newblock {\em PLOS ONE}, 9(11):e111506, 2014.

\bibitem{Thij14}
Marijn Thij, Tanneke Ouboter, Dani{\"e}l Worm, Nelly Litvak, Hans Berg, and
  Sandjai Bhulai.
\newblock {\em Algorithms and Models for the Web Graph: 11th International
  Workshop, WAW 2014, Beijing, China, December 17-18, 2014, Proceedings},
  chapter Modelling of Trends in Twitter Using Retweet Graph Dynamics, pages
  132--147.
\newblock Springer International Publishing, Cham, 2014.

\bibitem{Cavers78}
JK~Cavers.
\newblock On the {F}ast {F}ourier {T}ransform inversion of probability
  generating functions.
\newblock {\em IMA Journal of Applied Mathematics}, 22(3):275--282, 1978.

\bibitem{Marder07}
M~Marder.
\newblock Dynamics of epidemics on random networks.
\newblock {\em Physical Review E}, 75(6):066103, 2007.

\bibitem{Abate92}
Joseph Abate and Ward Whitt.
\newblock Numerical inversion of probability generating functions.
\newblock {\em Operations Research Letters}, 12(4):245--251, 1992.

\bibitem{Talbot79}
Alan Talbot.
\newblock The accurate numerical inversion of {L}aplace transforms.
\newblock {\em IMA Journal of Applied Mathematics}, 23(1):97--120, 1979.

\bibitem{Abate06}
Joseph Abate and Ward Whitt.
\newblock A unified framework for numerically inverting {L}aplace transforms.
\newblock {\em INFORMS Journal on Computing}, 18(4):408--421, 2006.

\bibitem{Abate04}
J~Abate and PP~Valk{\'o}.
\newblock Multi-precision {L}aplace transform inversion.
\newblock {\em International Journal for Numerical Methods in Engineering},
  60(5):979--993, 2004.

\bibitem{Myers14}
Seth~A Myers and Jure Leskovec.
\newblock The bursty dynamics of the {T}witter information network.
\newblock In {\em Proceedings of the 23rd International Conference on World
  Wide Web}, pages 913--924. ACM, 2014.

\end{thebibliography}
\bibliographystyle{unsrt}

\end{document}